\documentclass[12pt]{article}
\usepackage{epsfig,amssymb}
\usepackage{latexsym}

\newcommand{\bq}{\begin{equation}}
\newcommand{\eq}{\end{equation}}
\newcommand{\bqa}{\begin{eqnarray}}
\newcommand{\eqa}{\end{eqnarray}}
\newcommand{\ben}{\begin{enumerate}}
\newcommand{\een}{\end{enumerate}}
\newcommand{\bc}{\begin{center}}
\newcommand{\ec}{\end{center}}
\newcommand{\bqb}{\begin{eqnarray*}}
\newcommand{\eqb}{\end{eqnarray*}}

\def\swsq{s^2_W}

\def\mwsq{m_W^2}
\def\mzsq{m_Z^2}

\def\tchi{\tilde \chi}

\def\pr#1#2#3{ Phys. Rev. ${\bf{#1}}$, #2 (#3)}
\def\prl#1#2#3{ Phys. Rev. Lett. ${\bf{#1}}$, #2 (#3)}

\def\prep#1#2#3{ Phys. Rep. ${\bf{#1}}$, #2 (#3)}

\def\np#1#2#3{ Nucl. Phys. ${\bf{#1}}$, #2 (#3)}

\def\jhep#1#2#3{ JHEP ${\bf{#1}}$, #2 (#3)}
\def\epj#1#2#3{ Eur. Phys. J. ${\bf{#1}}$, #2 (#3)}
\def\ijmp#1#2#3{ Int. J. Mod. Phys. ${\bf{#1}}$, #2 (#3)}

\def\fortp#1#2#3{ Fortsch. Phys. ${\bf{#1}}$, #2 (#3)}

\def\polon#1#2#3{Acta Phys. Polon. ${\bf{#1}}$, #2 (#3) }

\begin{document}
\pagenumbering{arabic}
\thispagestyle{empty}
\def\thefootnote{\fnsymbol{footnote}}
\setcounter{footnote}{1}

\begin{flushright}
October  10,    2011.\\
arXiv:1106.2707 [hep-ph]\\
Corrected version.

 \end{flushright}

\vspace{2cm}

\begin{center}

{\Large {\bf Supersimplicity:}}\\
{\Large {\bf a remarkable high energy SUSY property.}}

 \vspace{1cm}
{\large G.J. Gounaris$^a$ and F.M. Renard$^b$}\\
\vspace{0.2cm}
$^a$Department of Theoretical Physics, Aristotle
University of Thessaloniki,\\
Gr-54124, Thessaloniki, Greece.\\
\vspace{0.2cm}
$^b$Laboratoire Univers et Particules de Montpellier,
UMR 5299\\
Universit\'{e} Montpellier II, Place Eug\`{e}ne Bataillon CC072\\
 F-34095 Montpellier Cedex 5.\\
\end{center}

\vspace*{1.cm}
\begin{center}
{\bf Abstract}
\end{center}

It is known that for any 2-to-2 process in MSSM,
only the helicity conserving
(HC) amplitudes survive asymptotically.   Studying many such processes,
  at the 1loop Electroweak (EW) order, it is found  that
  their high energy HC amplitudes are determined by just three forms: a log-squared function
  of the  ratio of two of the $(s,t,u)$ variables, to which  a $\pi^2$ is added;
  and two  Sudakov-like $\ln$- and  $\ln^2$-terms accompanied by respective
  mass-dependent  constants. Apart from a possible  additional
  residual constant (which is also discussed),
  these   HC amplitudes, may be   expressed
  as linear combinations of the above three forms, with  coefficients
  being  rational functions of the $(s,t,u)$ variables.
  This 1loop property, called  {\it supersimplicity}, is of course claimed
  for the 2-to-2 processes considered;
  but no violating examples  are known at present.
  For   $ug\to dW$, {\it supersimplicity} is found to
be a very good approximation  at LHC energies,
provided the SUSY scale is not too high. SM  processes are also discussed,
and their differences   are  explored.

\vspace{0.5cm}

\def\thefootnote{\arabic{footnote}}
\setcounter{footnote}{0}
\clearpage

\section{ Introduction}

Supersymmetry is well-known for its remarkable properties  controlling
the hierarchy problem and  improving  the realization of
 Grand Unification \cite{SUSYrev}. More recently, two additional properties
of Supersymmetry were noticed at the  high energy behavior
of the  scattering amplitudes, where the soft supersymmetry (SUSY) breaking effects
are minimized.

 The first one concerns the differences in the coefficients of
the 1loop electroweak (EW)
 logarithmic behaviors contained  in the so-called Sudakov terms, in  SM and MSSM
\cite{MSSMrules1, MSSMrules2, MSSMrules3, MSSMrules4}.
The second one refers to the  helicity conservation (HCns) property,
which is specific to Supersymmetry.

This HCns property  has been
 first proven to all orders in MSSM,
at the approximation where all soft SUSY breaking  effects,
as well as the $\mu$ bilinear term of the scalar sector, are neglected \cite{heli1, heli2}.
More explicitly it was  showed  that for any 2-to-2 processes
\bq
a_{\lambda_1}+b_{\lambda_2}\to c_{\lambda_3}+d_{\lambda_4} ~~, \label{gen-process}
\eq
where $\lambda_i$ denote the particle helicities, all amplitudes  violating
the helicity conservation rule
\bq
\lambda_1+\lambda_2=\lambda_3+\lambda_4 ~~, \label{HC-rule}
\eq
must  vanish at high  energies and fixed angles in MSSM \cite{heli1, heli2};
such amplitudes are called helicity violating (HV) amplitudes. Renormalizability is essential
for the validity of HCns; all known anomalous couplings violate it \cite{anomal}.

So, only the helicity conserving (HC) amplitudes obeying (\ref{HC-rule}), can survive
asymptotically in MSSM. But in \cite{heli1, heli2},
nothing was said  about the structure of the HC amplitudes at high energy,
where mass effects  may remain  important, at least so far as they affect the scale
of some logarithms. To study such mass effects in both,
 the HC and HV amplitudes, and investigate how  HCns is realized in MSSM
 and violated in SM,  many detail 1loop EW calculations have been performed.
 The main results  are summarized in the
 following paragraphs.\\

At the Born level, HCns is valid in  both, the SM and  the MSSM models.
In such a case all HV amplitudes vanish asymptotically like inverse powers of the energy,
while the HC ones tend to non-vanishing constants.
Particularly for processes involving external gauge bosons,
huge cancelations among the various diagrams contrive to establish HCns \cite{Corfu}.

At the 1loop EW level, with all mass terms retained,
the high energy helicity amplitudes have been investigated, in both SM and MSSM,
 for    gluon fusion producing a pair of gauge or Higgs bosons in \cite{ggVV, ggHH},
 and  for $ug\to dW$ in \cite{ugdW}.
In all MSSM cases, it has then been studied  how the high energy vanishing
of all HV amplitudes is realized; usually like an inverse power of the energy,
as the spartner contributions (sfermions and  inos)
cancel out the SM ones.  In SM, on the contrary, it is only accidentally
that the HV amplitudes may vanish asymptotically, and many cases have been identified where
this does not happen \cite{ggVV, ggHH}.\\

Concentrating on the HC asymptotic amplitudes in MSSM now, we  distinguish two
types of processes; those  where there is no Born terms, and the ones in which
Born terms are present. In each case, we  define  the
1loop EW order property of {\it supersimplicity},
 and explain how this definition  is modified as we go from MSSM to SM.\\

In the first case,
detail analytical  studies at 1loop EW order,
have recently  been done for the gluon fusion  to vector boson process $gg\to VV'$ \cite{ggVV},
and the chargino and  neutralino transitions  $gg \to \tchi_i\tchi_j$ \cite{ggchichi}.
In these cases, there are  no Born contributions
  and no Sudakov logarithms appear,  implying no dependance
on the SUSY breaking masses.

The HC asymptotic  structure is then solely  determined by forms like $\ln^2+ \pi^2$ ,
where   ratios of the $(s,t,u)$ Mandelstam variables appear within the quadratic logarithms.
The overall coefficients of these forms are solely determined by  rational
functions of  $(s,t,u)$, and there is no  additional term.
This is  the {\it supersimplicity} structure in this case.
All relevant formulae for this have already appeared in \cite{ggVV,ggchichi}, but they
were not related to the concept of  {\it supersimplicity}; this  we do here.\\

The more important and new work in the present paper, still within MSSM,
 concerns the second type of the above processes for which Born terms are present.
In this context we study the high energy 1loop EW amplitudes for the
2-to-2 processes,
\bq
u g\to d W ~,~ bg\to tW ~,~  bg\to tH^- ~,~ bg\to bZ ~,~
bg\to bH^0 ~,~ gg\to t\bar t ~,~ gg \to  \tilde t \tilde {\bar t} ~ ,
\label{process-Born}
\eq
 and  their SUSY transformed ones
\bq
\tilde u_L \tilde g\to \tilde d_L \tilde W ~,~
\tilde b \tilde g\to \tilde t \tilde W ~,~  \tilde b \tilde g\to \tilde t \tilde H^-
~,~ \tilde b \tilde g \to \tilde b \tilde Z ~,~
\tilde b \tilde g\to \tilde b \tilde H^0 ~,~
\tilde g \tilde g\to \tilde t \tilde{\bar t} ~,~
\tilde g \tilde g \to   t  \bar t ~ . \label{mirror-process-Born}
\eq
Note that  the processes  in (\ref{mirror-process-Born}) involve
    the gaugino and higgsino SUSY-counterparts of the
 charged and  neutral gauge and Higgs boson processes in (\ref{process-Born}).

For the processes (\ref{process-Born}, \ref{mirror-process-Born}),
the content of {\it supersimplicity} is  more involved.
 More explicitly, we find  that the  asymptotic HC amplitudes  are now
expressed as  linear combinations of three possible forms,  with coefficients being
rational functions of the $(s,~t,~u)$-variables.
The first of these forms is the  $\ln^2+ \pi^2$  one,  we have already seen for
 $gg\to VV',   \tchi_i\tchi_j$.
The other two forms  consist of two Sudakov like terms, involving log and
log-squared functions of a Mandelstam variable  scaled by masses,
to which   respective "constants" are  added,   depending on ratios of masses.

The constants entering the definition of these   three forms,  greatly enhance
the accuracy of the asymptotic expressions for the HC amplitudes,
 and allow to make valuable numerical predictions
for physical observables. In addition to these forms,
extra  "residual constants" may also appear
for the  on-shell renormalized amplitudes of the MSSM processes
(\ref{process-Born},\ref{mirror-process-Born}), at high energy.\\

 Thus, {\it supersimplicity} completes  the  previously known rules
 for the purely  logarithmic structure of  Sudakov
and angular depending  terms, determining
the high energy behavior of the 2-to-2 amplitudes \cite{MSSMrules2, MSSMrules3,MSSMrules4}.

While doing the analytical computations, we have also noticed an interesting
recipe for  obtaining  the high energy MSSM results.
This is based on  the remark  that  it is often easier to
first  compute the  relevant SUSY spartner process in
(\ref{mirror-process-Born}), and then obtain the result for the  actual process
in (\ref{process-Born}), through a SUSY transformation. This is because
the particles involved in the processes (\ref{mirror-process-Born}),
have usually smaller spins, than those in (\ref{process-Born}).\\

All together, the concept of {\it supersimplicity} in MSSM turns  out to have
three  aspects: the simplicity of the high energy HC amplitudes; the  recipe for
computing these expressions by using the SUSY transformed processes; and the
possibility of introducing  a very simple renormalization scheme, the supersimplicity
renormalization scheme (SRS), where only the above three forms appear asymtotically,
without any additional constant. This SRS scheme  may numerically be very close
to the on-shell scheme. At least, this is what we have seen for  $ug\to dW$,
where  the  supersimplicity structure may  be accurately (or approximately) valid
at LHC, provided the SUSY scale is in the TeV range (or just above it).\\

The purpose of the present work is to describe this
{\it supersimplicity} structure of  the high energy HC  amplitudes in MSSM,
and to study its numerical accuracy  for observable quantities. We repeat that this
property is only defined at the 1loop EW order.\\

Contents: Sect.2 summarizes the MSSM {\it supersimplicity} structure
of the  processes $gg \to VV'$, involving  no-Born term;
  based on  the results of   \cite{ggVV,ggchichi}.
In Sect.3, the  {\it supersimplicity} structure is described for    processes
(\ref{process-Born}, \ref{mirror-process-Born}), which contain a Born-contribution.
A detail study of $ug\to dW$ with numerical illustrations is also presented,
while an analogous discussion of  $bg\to b H_i^0$ appears in the Appendix.
The results of Section 3 and the Appendix, appear here for the first time.
Finally in Section 4, we present the Conclusions.\\

\section{Supersimplicity for  $gg\to VV'$ }

Here we summarize how {\it supersimplicity} appears in the 1loop EW order results
of \cite{ggVV, ggHH} for $gg\to VV'$,  where $V,V'$ are EW vector bosons.
The results cover not only the MSSM case, but the SM also.  \\

In MSSM, we of course have helicity conservation (HCns) at high energies.
The asymptotic HV amplitudes thus  vanish, while the HC ones are  expressed
through  the form
\bq
r_{xy}\equiv {-x-i\epsilon  \over -y-i\epsilon} ~~~~ \Rightarrow  ~~~~
\tilde d (r_{xy}) \equiv  \ln^2 r_{xy}+\pi^2 ~~,
\label{d-tilde}
\eq
 with $x$ and $y$ being any two of the $(s,t,u)$ Mandelstam variables.

For transverse  vector bosons, such high energy HC amplitudes
are given by \cite{gamgamZZ, ggVV}
\bq
F(gg\to ZZ)_{\mu\mu' \tau\tau'} =\alpha\alpha_s
{(9-18s^2_W+20s^4_W)\over 24 s^2_Wc^2_W} \delta_{\mu\mu' \tau\tau'}~~,
\label{Fasym-MSSM-ggZZ}
\eq
where $(\mu,\mu')$ denote the initial gluon helicities, while $(\tau,\tau')$
are the helicities of the final vector bosons, and
\bqa
& & \delta_{+-+-}=\delta_{-+-+}=-4 \tilde d (r_{ts})  ~~, \nonumber \\
& & \delta_{+--+}=\delta_{-++-}  = -4 \tilde d (r_{us}) ~~, \nonumber \\
&& \delta_{++++}  =\delta_{----}= -4  \tilde d (r_{tu})  ~~, \label{delta-MSSM-TT}
\eqa
while  all HV amplitudes satisfying $\mu+\mu'\neq \tau+\tau'$ vanish.
 A color factor $\delta^{ab}$, with  $(a,b)$
describing  the gluon $SU(3)$ indices,
is always removed from the amplitudes in (\ref{Fasym-MSSM-ggZZ}).  Similar expressions
for $gg\to \gamma\gamma, ~\gamma Z, ~W^+W^-$ exist also \cite{ggchichi}.

Thus, the {\it supersimplicity} structure in this  MSSM case means
that  all high energy  transverse HC amplitudes
are proportional\footnote{Real and Imaginary parts.} to the \underline{single}
form (\ref{d-tilde}), without any additional constant.

Contrarily to the  type of processes that we  study in Sect.3,
where additional forms  related to Sudakov logs appear; in the present
case, no such Sudakov terms arise. \\

The derivation of the 1loop asymptotic results (\ref{delta-MSSM-TT}, \ref{Fasym-MSSM-ggZZ})
from \cite{ggVV, gamgamZZ} is quite laborious. A much simpler way to obtain  them,
is by looking at the SUSY-transformed processes
\bq
\tilde g \tilde g \to \tilde B\tilde B,~~ \tilde W^{(3)}\tilde W^{(3)},~~
\tilde W^+\tilde W^- ~~ , \label{gaugino-pair}
\eq
remembering that the signs of the gaugino-helicities are the same as those
of the transverse gauge-bosons from which they were obtained,
through the SUSY-transformation
\cite{heli1, heli2}.
In such a case, the box diagrams involve only 2 fermionic lines,
each  with only one $\gamma^{\mu}$ matrix. The calculation is then much simpler,
leading, for transverse gauge bosons,  to \cite{ggchichi}
\bq
 (-1)^{\tilde \mu-\tilde \tau'}
F(\tilde g \tilde g \to \tilde V \tilde V')_{\tilde \mu \tilde \mu' \tilde\tau \tilde\tau'}
=F(gg \to VV')_{\mu\mu' \tau\tau'} ~~,
\label{gauge-gaugino-asym}
\eq
where $\tilde \mu, ~ \tilde \mu',~ \tilde\tau, ~ \tilde\tau' $
are the gluino and gaugino helicities, which of course receive  half integers values.
The r.h.s. of (\ref{gauge-gaugino-asym}) is of course determined by (\ref{Fasym-MSSM-ggZZ}),
and similar expressions for the other  gauge bosons. As seen in (\ref{gauge-gaugino-asym}),
 most of the gauge and gaugino asymptotic amplitudes,
are  identical. But for $\tilde \mu-\tilde \tau'=\pm 1$,
sign differences appear, related to the way the fermionic states in the l.h.s. of
 (\ref{gauge-gaugino-asym}) are defined.

An important role for  the validity of  (\ref{gauge-gaugino-asym}), is played by
the fact that the  asymptotic amplitudes for   (\ref{Fasym-MSSM-ggZZ},\ref{gaugino-pair})
are  mass-independent; this allows us to consider
un-mixed states. This is not true for the processes in Section 3, where
mass complications always  appear in the HC asymptotic 1loop amplitudes. \\

Results analogous to (\ref{gauge-gaugino-asym}), are also true
for longitudinal vector bosons, which necessarily include    higgsino
 amplitudes  in the l.h.s. \cite{ggchichi, ggVV, gamgamZZ}.
 Thus,  in order to study the $gg\to VV'$ asymptotic behavior,
it is advantageous  to  consider the SUSY-transformed
process $\tilde g \tilde g \to \tchi_i \tchi_j$,
with the appropriate gaugino and   higgsino  $\tchi_i \tchi_j$ components.
Such a   procedure     simplifies the calculation a lot\footnote{This way,
one  obtains  that the $gg\to VH$ processes
are mass suppressed, at high energy, because of the left-right orthogonality
of the gaugino-higgsino contributions.}. \\

The asymptotic   structure in SM   is  mutilated  by  additional
$A^S$ contributions,  inducing  non vanishing HV asymptotic amplitudes,
and at the same time also creating  HC  contributions which include forms other than
 (\ref{d-tilde}). Explicitly,  the  SM asymptotic amplitudes  for transverse
  final vector bosons are \cite{ggVV}
\bq
F(gg\to ZZ)^{SM}_{\mu\mu' \tau\tau'} =\alpha\alpha_s
{(9-18s^2_W+20s^4_W)\over 24 s^2_Wc^2_W}
[\delta_{\mu\mu' \tau\tau'}-2A^S_{\mu\mu' \tau\tau'}]~~, \label{Fasym-SM-ggZZ}
\eq
where
$\delta_{\mu\mu' \tau\tau'}$ only contributes to the HC amplitudes and is given
by (\ref{delta-MSSM-TT}); while  $A^S$ contributes,
both to the HC and HV transverse amplitudes as
\bqa
&& A^S_{++++}=A^S_{----}=4-{4ut\over s^2} \tilde d (r_{tu})
+{4(t-u)\over s}\ln \left ({t\over u} \right ) ~~, \nonumber \\
&& A^S_{+-+-}=A^S_{-+-+}=4-{4st\over u^2} \tilde d (r_{st})
+{4(s-t)\over u}  \ln \left ({-s-i \epsilon \over -t} \right ) ~~, \nonumber \\
&& A^S_{+--+}=A^S_{-++-}=4-{4su\over t^2} \tilde d (r_{su})
+{4(s-u)\over t}  \ln \left ({-s-i \epsilon \over -u} \right )~~, \label{ASHC}
\eqa
and
\bqa
&&A^S_{+++-}=A^S_{+-++}=A^S_{++--}=A^S_{++-+}=A^S_{+---}=A^S_{---+}=A^S_{-+--}
\nonumber\\
&& =A^S_{--++}=A^S_{--+-}=A^S_{-+++}~=~-4  ~~. \label{ASHV}
\eqa

In all these SM cases, receiving no  Born contribution,
the asymptotic   HV amplitudes behave like constants.
On the contrary,  the high energy HC
 amplitudes are linear combinations of the form  (\ref{d-tilde})
and   single   logarithms of ratios of the $(s,t,u)$ variables;
to which   additional constants,  like those in (\ref{ASHC}) are added.
Thus, the {\it supersimplicity} structure is somewhat reduced in SM.

Such linear logarithmic  and additional constant terms, are never  seen
in the    MSSM case (\ref{Fasym-MSSM-ggZZ}) \cite{ggVV, ggHH}.\\

The asymptotic HC amplitudes involving longitudinal $ZZ$ and $W^+W^-$ final states,
in both SM and MSSM,
 are  solely   determined by the single logaritmic form (\ref{d-tilde}),
with coefficients being rational functions of the Mandelstam variables; see eqs.(25) of
\cite{ggVV}. \\

\section{ Processes with Born terms at high energies.}

We here consider the high energy behavior of the  processes
(\ref{process-Born}, \ref{mirror-process-Born}), which receive
non-vanishing  Born contributions; the results thus obtained have not appeared in
previous publications. According to these,   the high energy behavior of
the 1loop HC amplitudes is determined by the form (\ref{d-tilde}),
as those of the previous section; but in addition to it, two new forms appear,
containing   the so-called Sudakov $\ln^2$ and $\ln$ terms  \cite{Sud}, to which
specific "constant" corrections are added.

The coefficients of $\ln^2$ are known
to be identical in MSSM and SM, while those of the linear-$\ln$
terms are  clearly different, even when disregarding
the mass-scales inside logarithms  \cite{MSSMrules1, MSSMrules2, MSSMrules3}.

The emphasis  here though, is on  the aforementioned  "constant" corrections,
which accompany   the  logarithms and depend on  ratios of masses,
in both, MSSM and SM.  The "augmented Sudakov logarithms" thus  introduced in Section 3.1,
considerably enhance the  accuracy of the high energy expressions.\\

In  MSSM, the only asymptotically non-vanishing  amplitudes for the processes
(\ref{process-Born}, \ref{mirror-process-Born}) at high energy,
 are the HC ones \cite{heli1, heli2}.
 At the 1loop EW order,  a simple correspondence between the amplitudes of
(\ref{process-Born}) and those of (\ref{mirror-process-Born}) has been found.
This  is not an exact equality,
like in (\ref{gauge-gaugino-asym}); but an equivalence    of the forms,
which are of course mass dependent. Thus, the results for (\ref{process-Born}), may be simply
obtained by renaming those of the corresponding process in (\ref{mirror-process-Born}).
The external and internal masses of the process (\ref{mirror-process-Born})
have just to be replaced by the ones
of\footnote{ This is facilitated when chargino, neutralino
and squark mixings are neglected.} (\ref{process-Born}).

As the complexity of the calculation  increases with the spin
of the particles involved,  the computation of the
processes (\ref{mirror-process-Born}), is usually   much simpler
than those of the processes in (\ref{process-Born}).
Thus, it is often advantageous to first calculate the HC amplitudes in
the interesting SUSY-transformed process
of type (\ref{mirror-process-Born}), and then   translate the result to
the one for the original process in  (\ref{process-Born}).

We next turn to the    augmented Sudakov logarithms, mentioned above.\\

\subsection{The augmented Sudakov forms and Supersimplicity.}

For any 2-to-2 processes, at 1loop EW order, in either MSSM or SM,
 there are two  augmented Sudakov forms; the form  $\ln^2$ and the form $\ln$.
The $\ln^2$-form  is generated  completely  from each contributing diagram;
i.e. they are not the result of combining contributions from different diagrams.
This is also true for the form  in (\ref{d-tilde}).
In contrast, for the linear $\ln$-form,
different diagrams, including self-energy contributions, conspire to  generate it;
this happens in the same way the divergent parts cancel.

In both cases, the Sudakov  logs are accompanied by   dimensionless
"constants" depending
  on one of  the external masses of the considered process, and two internal  masses
  of the generating diagrams.   These diagrams  of course  contain  a vertex  where the
  two   internal lines join to produce the   external one. \\

 The augmented Sudakov  $\ln^2$-form is   generated by triangular or box-diagrams
 with gauge boson exchanges,
 and it    involves the logarithm-squared
 of a Mandelstam  $(s,t,u)$ variable  scaled  by  a gauge boson
 mass\footnote{It is conceivable that other masses of internally exchanged particles
 may also affect  this scale;  e.g. a Higgs mass.}, in all examples we know
 \cite{MSSMrules1, MSSMrules2,MSSMrules3}.  Its   general structure  is
 \bq
\overline{\ln^2s_V}\equiv
\ln^2\left (\frac{-s-i\epsilon}{m_V^2} \right )+2L_{a_1Vc_1}+2L_{a_2Vc_2} ~~,
\label{Sud-ln2}
\eq
 and  similarly for the  $t,u$ variables\footnote{For $V=W$, the notation $s_W$
 in (\ref{Sud-ln2}), should not be confused by the coincidence with the notation for the
 sine of the Weinberg-angle.}.   Here\footnote{To regularize infrared singularities we use
  $m_\gamma=m_Z$. The same choice was made in  \cite{ugdW}.} $m_V=m_W,m_Z,m_\gamma$.
  The  constant term in the r.h.s. of  (\ref{Sud-ln2}) is given by
  \cite{Denner1, Denner2, asPV}.
 \bqa
L_{aVc}\equiv L(p_a, m_V, m_c) & = &
 \phantom{+} {\rm Li_2} \left ( \frac{2p_a^2+i\epsilon}{m_V^2-m_c^2+p_a^2+i\epsilon +
\sqrt{\lambda (p_a^2+i\epsilon, m_V^2, m_c^2)}} \right )
\nonumber \\
&& + {\rm Li_2} \left ( \frac{2p_a^2+i\epsilon }{m_V^2-m_c^2+p_a^2+i\epsilon -
\sqrt{\lambda (p_a^2+i\epsilon, m_V^2, m_c^2)}} \right )~~,
\label{LaVc-term}
\eqa
where ${\rm Li_2}$ is a Spence function and
\bq
\lambda(a,b,c)=a^2+b^2+c^2-2ab-2ac-2bc~~. \label{lambda-function}
\eq
The complex quantities   $L_{aVc}$ of (\ref{LaVc-term}), are ubiquitous
in the asymptotic expansion of the
Passarino-Veltman (PV) functions  \cite{PV, asPV}.
The first index in them refers to an external particle $(a)$
of  the considered  processes, with its mass and momentum  satisfying
$p_a^2=m_a^2$; while the other two indices describe the masses
 $(m_V, m_c)$  of two internal  particles $(V,c)$ in the generating diagram,
 joining to  the $aVc$-vertex.  Since any $V$ internal line
has two ends,  there are always two such terms generated by each contributing
 diagram, called  $L_{a_1Vc_1}$ and $L_{a_2Vc_2}$,
 which lead to the two\footnote{If $c_1$ or $c_2$,
is actually a mixed state of several particles,
then all of them will appear in (\ref{Sud-ln2}),
increasing the number of terms in it.} last terms in  (\ref{Sud-ln2}).\\

We next turn to the augmented Sudakov $\ln$ forms, generated by
bubble\footnote{Relevant for self-energy and counter term contributions.},
triangular or box diagrams.  These diagrams always involve two internal
lines  $(i,j)$, joining to a vertex where an external particle
$(a)$ is produced, through a non vanishing $(ija)$-coupling.
Its general form is
\bq
\overline{\ln s_{ij}}\equiv \ln \frac{-s-i\epsilon}{m_im_j}+b^{ij}_0(m_a^2)-2 ~~,
\label{Sud-ln}
\eq
and  similarly for the  $t,u$ variables. Here $b_0^{ij}(m_a^2)$ is a
finite part of the standard  $B_0$ bubble-function, defined as \cite{PV, asPV}
\bqa
b_0^{ij}(m_a^2)& \equiv& b_0(m_a^2; m_i,m_j) =
2 + \frac{1}{m_a^2} \Big [ (m_j^2 -m_i^2)\ln\frac{m_i}{m_j}\nonumber\\
&& + \sqrt{\lambda(m_a^2+i\epsilon, m_i^2, m_j^2)}  {\rm ArcCosh} \Big
(\frac{m_i^2+m_j^2-m_a^2-i\epsilon}{2 m_i m_j} \Big ) \Big ] ~~. \label{b0ij}
\eqa  \\

In MSSM, the content of {\it supersimplicity}
for the  Born-containing processes (\ref{process-Born}, \ref{mirror-process-Born}),
 is the following.
 First,  a {\it supersimplicity}  renormalization scheme (SRS) may be defined
for these  process,
where the asymptotic HC amplitudes only contain   linear combinations
of the above three forms, with coefficients being rational functions of the
$(s,t,u)$ variables. Then, the high energy  HC amplitudes,
in the usual on-shell (OS) scheme \cite{OS}, may be  \underline{completely} expressed
as the addition of the aforementioned SRS amplitudes, to which a
"residual constant" is added. This
"residual constant" then  acts as  a counter term relating
the SRS and  on-shell schemes, and it may be
very small; at least this is what we have found in the $ug\to dW$ case of Sect. 3.2.\\

What happens in SM? In this case, helicity conservation
is not valid  to all orders; but it holds at the Born level, for any 2-to-2 process.
 Because of this, for  Born-involving  processes,  the high energy HV  amplitudes
 are usually  much smaller than the HC ones. This is also true for
  $\gamma \gamma \to \gamma \gamma$,  $\gamma Z, ZZ$
  \cite{gamgamgamgam, gamgamgamZ, gamgamZZ}; but not for  $gg\to VV'$
\cite{ ggVV, ggHH}.

Concentrating on the  HC amplitudes,  and restricting to the Born-involving
processes (\ref{process-Born}), we find that the high energy structure
in this SM case
may again be described by  the forms
(\ref{d-tilde},\ref{Sud-ln2},\ref{Sud-ln}), (with different coefficients of course),
but this time an additional form also appears
 involving   linear logarithms of  ratios of any two of the  $(s,t,u)$-variables;
 i.e. there are four different forms in the SM case. In addition to them though,
 "SM residual constants" are needed to describe the on-shell
 amplitudes.

Again, a renormalization scheme, in analogy to  SRS,
may  be defined for  SM,  where all asymptotic HC amplitudes
are expressed as linear combinations of the aforementioned
four forms, without any additional residual constant.

Below we call this scheme also SRS, in spite of the fact that
we now refer to SM and not to MSSM. Again, the aforementioned "SM residual constants"
act as counter terms  relating  SRS and to the on-shell scheme.\\

In the next Sect.3.2 we  give illustrations of
the asymptotic HC  amplitudes for $ug\to dW$,
in both MSSM and SM. Corresponding results for $bg\to b H^0_i$,  appear
in the Appendix. \\

Before finishing this Section we also add a remark on
the no-Born processes $\gamma \gamma \to \gamma \gamma$, $\gamma Z, ZZ$,
for which,  of course,
no renormalization scheme dependence arises.
In such a case, the high energy HC amplitudes in MSSM
only contain the forms (\ref{d-tilde}, \ref{Sud-ln2})
\cite{gamgamgamgam, gamgamgamZ, gamgamZZ}. In SM though, the asymptotic HC amplitudes
 contain also   linear logarithms involving rations of two of the $(s,t,u)$ variables,
  as well as     additional  constant terms.  In both cases the contribution of the form
  (\ref{Sud-ln2}) is induced by the $W$-loop .\\

\subsection{ High energy $ug\to dW$  amplitudes at 1loop EW order.}

In order to appreciate the usefulness and  accuracy of the
{\it supersimplicity} description,
we here present analytical expressions for
 the high energy   HC amplitudes of the process
$ug\to dW$, to the  1loop EW order. Previous semi-analytical  results for these have
appeared in \cite{ugsdWino};  but there, the numerical components  were blurring the
picture and  the {\it supersimplicity} structure was  not visible.

We choose this process, because
its external particles are rather light,
 so that the asymptotic region may be   approached  quickly,
provided the SUSY scale is not too high. Moreover, since these external
particles exist already in SM, the analysis may be done, both in MSSM and SM.
This will be helpful in clarifying the SM-MSSM differences.\\

The complete EW 1loop helicity amplitudes have already been computed,
in the on-shell renormalization scheme \cite{OS}, in both MSSM and SM
  \cite{ugdW}. Denoting the $ug\to dW$ helicity amplitudes
  as\footnote{The indices describe respectively the $u$, $g$, $d$ and $W$ helicities.}
  $F_{\lambda \mu \tau \mu'}$,  the  two
  independent HC amplitudes are    $F_{-+-+}$ and $F_{----}$.

At high energies,  the on-shell (OS) HC amplitudes   may be written as
\bq
F^{\rm OS }_{-\pm-\pm} = F^{\rm Born}_{-\pm-\pm}\left [1+{\alpha\over4\pi}(C_{-\pm-\pm}
+\delta C_{\rm residual})\right ]~~~, \label{F-asym-ugdw}
\eq
where
\bq
F^{\rm Born}_{-+-+}={eg_s\over\sqrt{2}s_W}\left (2\cos{\theta\over2} \right )
~~~,~~~
F^{\rm Born}_{----}={eg_s\over\sqrt{2}s_W}\left ({2\over\cos{\theta\over2}} \right)
~~, \label{Born-asym-ugdw}
\eq
describe their asymptotic  Born expressions.  A color matrix factor $\lambda^a/2$,
acting between the initial $u$ and final $d$ quarks,
is always removed from  (\ref{F-asym-ugdw},\ref{Born-asym-ugdw}).
The {\it supersimplicity} structure
is contained in $C_{-\pm-\pm}$,  while
  $\delta C_{\rm residual}$ denotes  the residual constant  correction needed in the
  on-shell scheme.  \\

In MSSM, the results for  $C_{-\pm-\pm}$, may be computed in 2 different manners. Either
through a lengthy direct computation of the  $ug\to dW$ diagrams; or
in a much simpler way, by looking at the SUSY transformed process
 $\tilde u_L \tilde g\to \tilde d_L \tilde W$. In both cases of course, the asymptotic
 limit of the PV functions given in \cite{asPV} is used, which suffices for determining the
 high energy 2-to-2  physical amplitudes, up to energy suppressed terms.

In the first manner based on the complete  $ug\to dW$ results \cite{ugdW},
the only needed  augmented Sudakov $\ln^2$
forms of type (\ref{Sud-ln2}),  are
\bqa
&& \overline{\ln^2 t_Z}\equiv  \ln^2\frac{-t-i\epsilon}{\mzsq}  +2(L_{dZd}+L_{uZu})~~,
\nonumber \\
&& \overline{\ln^2 u_Z} \equiv  \ln^2\frac{-u-i\epsilon}{\mzsq}+2(L_{WZW}+L_{uZu})~~, \nonumber \\
&&  \overline{\ln^2 u_W}\equiv  \ln^2\frac{-u-i\epsilon}{\mwsq} +2(L_{WWZ}+L_{uWd})~~ ,\nonumber \\
&& \overline{\ln^2 s_Z} \equiv  \ln^2{-s-i\epsilon\over \mzsq} +2(L_{dZd}+L_{WZW})~~, \nonumber \\
&& \overline{\ln^2 s_W} \equiv  \ln^2{-s-i\epsilon\over m^2_W} +2(L_{dWu}+L_{WWZ}) ~~;
\label{Sud-ln2-ugdW}
\eqa
while for the augmented Sudakov $\ln$ forms of type (\ref{Sud-ln}), the relevant internal particles
$ij$ are such that either $ij=qV$ with $(V=W,~Z,~\gamma)$ and $(q=u,d)$, or
  $ij=\tilde q_L \tchi_j$ with $\tchi_j$ being a chargino or neutralino
   and $(\tilde q_L= \tilde u_L,\tilde d_L)$; leading to  respective quantities
like  $b_0^{uW}(m_d^2)$,  $b_0^{uZ}(m_u^2)$  or $b_0^{\tilde u_L\tchi_j^+}(m_d^2)$,
$b_0^{\tilde u_L\tchi_j^0}(m_u^2)$ etc.   \\

 Using then (\ref{F-asym-ugdw}), the complete 1loop EW results for $ug\to dW$ \cite{ugdW},
  lead to
 \bqa
&& C_{-+-+}^{\rm MSSM}= {(1-10c^2_W) \over 36c^2_Ws^2_W}\Bigg [-\overline{\ln^2 t_Z}
  - {t\over u}\Big (\ln^2 r_{ts}+\pi^2\Big )
  +\ln^2r_{tu}+\pi^2 \Bigg ]
\nonumber \\
&& +{1\over2s^2_W}\Bigg [-\overline{\ln^2 u_Z} - \overline{\ln^2 u_W}
  - \overline{\ln^2 s_Z} - \overline{\ln^2 s_W}
  +2\Big (\ln^2 r_{us}+\pi^2\Big ) \Bigg ]
\nonumber\\
&& +{(1+8c^2_W)\over24c^2_Ws^2_W}\Big [\overline{\ln s_{uZ}}
+\overline{\ln s_{d Z }}   \Big ]
 +{3\over4s^2_W}\Big [\overline{\ln s_{dW}}+
 \overline{\ln s_{uW}}   \Big ]
\nonumber\\
&&
-\sum_i\Bigg \{ {|Z^N_{1i}s_W +3 Z^N_{2i}c_W|^2\over72c^2_Ws^2_W}
\overline{\ln s_{\tilde u_L  \tchi^0_i }}
 + {|Z^N_{1i}s_W -3 Z^N_{2i}c_W|^2\over72c^2_Ws^2_W}
\overline{\ln s_{ \tilde d_L  \tchi^0_i }}
\nonumber \\
&& +  {|Z^-_{1i}|^2\over 4s^2_W} \overline{\ln s_{ \tilde d_L  \tchi^+_i }}
  + {|Z^+_{1i}|^2\over 4s^2_W} \overline {\ln s_{ \tilde u_L  \tchi^+_i}}
 \Bigg \}  ~~, \label{MSSM-Cmpmp} \\
&& C_{----}^{\rm MSSM}= {(1-10c^2_W) \over 36c^2_Ws^2_W}\Bigg [-\overline{\ln^2 t_Z}
  - {t\over s}\Big (\ln^2 r_{tu}+\pi^2\Big )
  +\ln^2 r_{ts}+\pi^2 \Bigg ]
\nonumber \\
&& +{1\over2s^2_W}\Bigg [-\overline{\ln^2 u_Z} - \overline{\ln^2 u_W}
  - \overline{\ln^2 s_Z} - \overline{\ln^2 s_W}
  +2\Big (\ln^2 r_{us}+\pi^2\Big ) \Bigg ]
\nonumber\\
&& +{(1+8c^2_W)\over24c^2_Ws^2_W}\Big [\overline{\ln u_{uZ}}
+\overline{\ln u_{d Z }}   \Big ]
 +{3\over4s^2_W}\Big [\overline{\ln u_{dW}}+  \overline{\ln u_{uW}}   \Big ]
\nonumber\\
&&
-\sum_i\Bigg \{ {|Z^N_{1i}s_W +3 Z^N_{2i}c_W|^2\over72c^2_Ws^2_W}
\overline{\ln u_{\tilde u_L  \tchi^0_i }}
 + {|Z^N_{1i}s_W -3 Z^N_{2i}c_W|^2\over72c^2_Ws^2_W}
\overline{\ln u_{ \tilde d_L  \tchi^0_i }}
\nonumber \\
&& +  {|Z^-_{1i}|^2\over 4s^2_W} \overline{\ln u_{ \tilde d_L  \tchi^+_i }}
  + {|Z^+_{1i}|^2\over 4s^2_W} \overline {\ln u_{ \tilde u_L  \tchi^+_i}}
 \Bigg \}  ~~, \label{MSSM-Cmmmm}
\eqa
which indeed contain only the forms\footnote{Note that $(C_{-+-+},~C_{----})$
are related to each-other through an $s \leftrightarrow u$
interchange. The same is true for the SM results in (\ref{SM-Cmpmp}, \ref{SM-Cmmmm}).}
 (\ref{d-tilde}, \ref{Sud-ln2},  \ref{Sud-ln}). The coefficients  $Z^N$ and $Z^+,~Z^-$
 in (\ref{MSSM-Cmpmp}, \ref{MSSM-Cmmmm}) describe the neutralino and
 chargino mixing matrices respectively \cite{Rosiek}.

The high energy HC amplitudes in the SRS scheme, simply become
\bq
F^{\rm SRS}_{-\pm-\pm} = F^{\rm Born}_{-\pm-\pm}\left [1+{\alpha\over4\pi} C_{-\pm-\pm}
\right ]~~~. \label{F-asym-ugdw-SRS}
\eq
Substituting in it,  the MSSM result (\ref{MSSM-Cmpmp}, \ref{MSSM-Cmmmm}),
we obtain the high energy MSSM HC amplitudes in the SRS scheme.\\

We next  discuss the additional "residual" contribution needed for calculating the
  on-shell (OS) result; compare   (\ref{F-asym-ugdw}) and note that
  the on-shell scheme has also been used in  the exact 1loop calculation of \cite{ugdW}.
 The "residual" contribution in (\ref{F-asym-ugdw})  arises from the $u$- and $d$-quark
 wave function renormalization constants   \cite{eeV1V2-long}
\bq
\delta Z^q_L={\alpha\over4\pi}\left [-c^{ij}_q(\Delta-ln{m_im_j\over\mu^2}+b^{ij}_0) \right ]
+\overline{\delta Z^q_L} ~~,  \label{delta-res-q}
\eq
with $\Delta$ being the usual divergent contribution and
 $c^{ij}_q$ is   the coupling coefficient for the $b_0^{ij}$ bubble (\ref{b0ij}).
 The  W field renormalization constants are  \cite{eeV1V2-long}
\bqa
&& \delta Z^W_1-\delta Z^W_2+{1\over2}\delta \Psi_W\equiv {\alpha\over4\pi}
\left [-~{2\Delta\over s^2_W}
+{1\over s^2_W}(2ln{m_Zm_W\over\mu^2}-2b^{ZW}_0)\right ]+\overline{\delta_W} ~~, \label{rctW} \\
&& \delta\Psi_W=-Re\hat\Sigma^{T'}_{WW}(m^2_W)=
-\{Re\Sigma^{T'}_{WW}(m^2_W) + \delta Z_W \}~~. \label{delta-psiW}
\eqa

Ignoring the square bracket  parts in (\ref{delta-res-q},\ref{rctW}),
that are already contained in the  supersimplicity $C_{-\pm-\pm}$-results
(\ref{MSSM-Cmpmp}, \ref{MSSM-Cmmmm}),  the actual residual correction
 in (\ref{F-asym-ugdw}),  may be written as
\bqa
\delta_{OS} & \equiv & {\alpha \over 4\pi} \delta C_{\rm residual}=
{1\over2}[\overline{\delta Z^u_L}+\overline{\delta Z^d_L}]
+\overline{\delta_W} \nonumber \\
&=& \frac{\alpha }{2\pi \swsq}\Big [ \ln\frac{m_W}{m_Z}+b_0^{ZW}(\mwsq) \Big ]
\nonumber \\
&& +{1  \over 2 } \left [\delta\Psi_W + ( \delta Z^d_L +\delta Z^u_L)
-(\delta Z^d_L +\delta Z^u_L)_{(B_1\to -B_0/2)}\right ]~~~,  \label{residual-COS}
\eqa
where   $(B_0,~B_1)$ are the standard PV bubble functions \cite{PV}.\\

In MSSM, the supersimplicity  expressions (\ref{MSSM-Cmpmp}, \ref{MSSM-Cmmmm}) may also
  be obtained in a much simpler  way,
by considering the process
 $\tilde u_L \tilde g\to \tilde d_L \tilde W$.
 The HC asymptotic amplitudes in this case are determined in terms of
 $(C_{++},~ C_{--})$ defined
 in analogy to (\ref{F-asym-ugdw}), with their indices describing the $\tilde g, ~ \tilde W$
 helicities. In this case, the first and third indices in the Sudakov $\ln^2$ forms
  $L_{aVc}$  of  (\ref{Sud-ln2-ugdW}) are changed to $L_{\tilde a V \tilde c}$; while
the linear  Sudakov $\ln$ forms  defined in (\ref{Sud-ln}) acquire   constant contributions
like\footnote{Note that the $b_0^{ij}$ functions in
the $\tilde u_L \tilde g\to \tilde d_L \tilde W$ are calculated at squark-masses,
as opposed to the $ug\to dW$ case, where they are calculated at the much smaller values
of the  $u$-  and $d$-quark masses. }  $b_0^{W\tilde u_L}(m_{\tilde d}^2)$ or
$b_0^{\tilde W u }(m_{\tilde d}^2)$ etc.
Transforming  $(C_{++},~ C_{--})$ back to the $ug\to dW$ case, we recover exactly
(\ref{MSSM-Cmpmp}, \ref{MSSM-Cmmmm}).\\

In SM, no SUSY transformation trick is applicable. In order to get the SM
high energy  amplitudes, we  have to work with the complete 1loop results of  \cite{ugdW},
suppressing the  SUSY exchange diagrams. Ignoring also the small high energy HV amplitudes,
and using for the HC ones the same   definitions
(\ref{F-asym-ugdw}, \ref{F-asym-ugdw-SRS}), we get
\bqa
&& C_{-+-+}^{\rm SM}= {(1-10c^2_W) \over 36c^2_Ws^2_W}\Bigg [-\overline{\ln^2 t_Z}
  + {t^2\over u^2}\Big (\ln^2r_{ts}+\pi^2\Big )
  +\ln^2r_{tu}+\pi^2  - \frac{2s}{u} \ln r_{ts}   \Bigg ]
  \nonumber \\
 &&   +{1\over2s^2_W}\Bigg [-\overline{\ln^2 u_Z} - \overline{\ln^2 u_W}
  - \overline{\ln^2 s_Z} - \overline{\ln^2 s_W}
  +2\Big (\ln^2r_{us}+\pi^2\Big ) \Bigg ]
\nonumber\\
&& +{(1+8c^2_W)\over24c^2_Ws^2_W}\Big [\overline{\ln s_{uZ}}
+\overline{\ln s_{d Z }}   \Big ]
 +{3\over4s^2_W}\Big [\overline{\ln s_{dW}}+
 \overline{\ln s_{uW}}   \Big ] ~~, \label{SM-Cmpmp} \\
&& C_{----}^{\rm SM}= {(1-10c^2_W) \over 36c^2_Ws^2_W}\Bigg [-\overline{\ln^2 t_Z}
  + {t^2\over s^2}\Big (\ln^2r_{tu}+\pi^2\Big )   +\ln^2r_{ts}+\pi^2
   - \frac{2u}{s} \ln r_{tu}   \Bigg ]
 \nonumber \\
 && +{1\over2s^2_W}\Bigg [-\overline{\ln^2 u_Z} - \overline{\ln^2 u_W}
  - \overline{\ln^2 s_Z} - \overline{\ln^2 s_W}
  +2\Big (\ln^2r_{us}+\pi^2\Big ) \Bigg ]
\nonumber\\
&& +{(1+8c^2_W)\over24c^2_Ws^2_W}\Big [\overline{\ln u_{uZ}}
+\overline{\ln u_{d Z }}   \Big ]
 +{3\over4s^2_W}\Big [\overline{\ln u_{dW}}+
 \overline{\ln u_{uW}}   \Big ] ~~,  \label{SM-Cmmmm}
\eqa
expressed completely in terms of the forms (\ref{d-tilde}, \ref{Sud-ln2},  \ref{Sud-ln})  and
linear logarithms $(\ln r_{ts}, \ln r_{tu})$.

Thus, using   (\ref{SM-Cmpmp}, \ref{SM-Cmmmm}) in  (\ref{F-asym-ugdw-SRS}),
we obtain  the SM asymptotic  HC amplitudes, in the SRS scheme.
Correspondingly, the residual correction needed in  the on-shell result
(\ref{F-asym-ugdw}) is again given by (\ref{residual-COS}),
where   the SUSY contributions
are now of course suppressed. Using (\ref{F-asym-ugdw}), together with
(\ref{SM-Cmpmp}, \ref{SM-Cmmmm}), the on-shell asymptotic SM amplitudes are obtained. \\

Starting from (\ref{F-asym-ugdw-SRS}, \ref{F-asym-ugdw}), the high energy  MSSM or SM amplitudes
in the   SRS and  OS  schemes are related by
\bq
F^{\rm OS}_{-\pm-\pm} = F^{\rm SRS}_{-\pm-\pm}\left [1+{\alpha\over4\pi}
\delta C_{\rm residual}
\right ]~~~, \label{F-OS-SRS}
\eq
leading to the definition of their percentage difference as
\bq
\delta_{OS} \equiv {\alpha \over 4\pi} \delta C_{\rm residual}=
\frac{F^{\rm OS}_{-\pm-\pm}-F^{\rm SRS}_{-\pm-\pm}}{F^{\rm SRS}_{-\pm-\pm}}~~.
\label{delta-OS-SRS}
\eq
Note that (\ref{F-OS-SRS},\ref{delta-OS-SRS}) clearly indicate that the real quantity
$\delta_{OS}$ acts like
a residual counter term relating the SRS and OS schemes.

We repeat that (\ref{F-asym-ugdw},\ref{F-asym-ugdw-SRS},\ref{F-OS-SRS},\ref{delta-OS-SRS})
are  valid in both, SM and MSSM, provided of course,
that the appropriate  $C_{-\mp-\mp}$ and $\delta C_{\rm residual}$ are used.

\begin{table}[h]
\begin{center}
{ Table 1: Input  parameters at the grand scale, for some   cMSSM  models\\
 with   $\mu>0$, and the  $ \delta_{OS}$ results.
 All dimensional parameters  in GeV. }\\
  \vspace*{0.3cm}
\begin{small}
\begin{tabular}{||c|c|c|c|c||c||}
\hline \hline
  & $m_{1/2}$ & $m_0$  &  $A_0$ & $\tan\beta$ & $ \delta_{OS}$     \\ \hline
$SPS1a'$ \cite{SPA1}  & 250   & 70& -300 & 10  & 0.0286   \\
 mSP4 \cite{mSP4}  & 137   & 1674& 1985  & 18.6  & 0.0292  \\
BBSSW \cite{Baer1}  & 900&  4716  & 0 & 30  & 0.0299  \\
BKPU \cite{Baer4} & 2900&  8700  & 0 & 50  & 0.0298  \\
\hline
ATLAS SU1  \cite{ATLAS-bench}  & 350 &  70 & 0 & 10  & 0.0289  \\
ATLAS SU2    & 300&  3550  & 0 & 10  & 0.0297  \\
ATLAS SU3    &300 &  100  & -300 & 6  & 0.0288  \\
ATLAS SU4    &160 &  200  & -400 & 10  & 0.0283  \\
ATLAS SU6    & 375 &  320  & 0 & 50  &  0.0290 \\
ATLAS SU8.1    & 360 &  210  & 0 & 40  & 0.0290  \\
ATLAS SU9    & 425 &  300  & 20 & 20  &  0.0291 \\
 \hline \hline
\end{tabular}
 \end{small}
\end{center}
\end{table}

To   compare the high energy  MSSM and SM predictions for the\footnote{For $ug\to dW$
above  0.5TeV, the HV  amplitudes are much smaller than the HC ones, in
all  MSSM benchmarks of  Table 1, and in SM \cite{ugdW}. }
 HC amplitudes  in the SRS scheme,
 we simply need to identify the differences between  (\ref{MSSM-Cmpmp}, \ref{MSSM-Cmmmm}) and
 (\ref{SM-Cmpmp}, \ref{SM-Cmmmm}).
Such differences    appear in the coefficients of the forms of type (\ref{d-tilde}) and
  (\ref{Sud-ln});  and most strikingly, in the SM linear
 logarithms of  ratios of the  $(s,t,u)$ variables, that  never appear  in MSSM.

 Constant "residual" contributions to the high energy HC amplitudes, in either    MSSM or  SM,
 (beyond those entering the aforementioned log-involving forms),
  can never appear  in the SRS scheme. They can appear in the on-shell scheme given
  by (\ref{F-asym-ugdw}) though,
 due to   the residual counter term   (\ref{delta-OS-SRS}), determined
 by  (\ref{residual-COS}).  \\

Coming now to the magnitude of the counter term  $\delta_{OS}$, relating
the OS and SRS schemes, we find from (\ref{residual-COS}) the
 numerical value
\bq
\delta_{OS}= {\alpha\over4\pi} \delta C_{\rm residual} \simeq 0.0289 ~~
\label{residual-SM-num}
\eq
in the SM case; while the results for a wide class of MSSM benchmarks,
are shown in the last column of Table 1.
The lower part of this table covers all ATLAS benchmarks
of  \cite{ATLAS-bench}, while the upper part covers also possibilities of very heavy
squarks and sleptons \cite{mSP4, Baer1, Baer4}.
 The counter terms $\delta_{OS}$ appear rather insensitive to the model,
the differences being at the unobservable permil level.

Consequently, the  {\it supersimplicity  SRS} amplitudes
and cross sections from (\ref{F-asym-ugdw-SRS}),
very closely approximate the on-shell ones from (\ref{F-asym-ugdw}).
Because of this, only the on-shell asymptotic results are plotted in the Figures,
where we  compare them to  the  complete one loop results
 in the same   scheme  \cite{ugdW}.\\

\begin{figure}[p]
\vspace*{-1cm}
\[
\epsfig{file=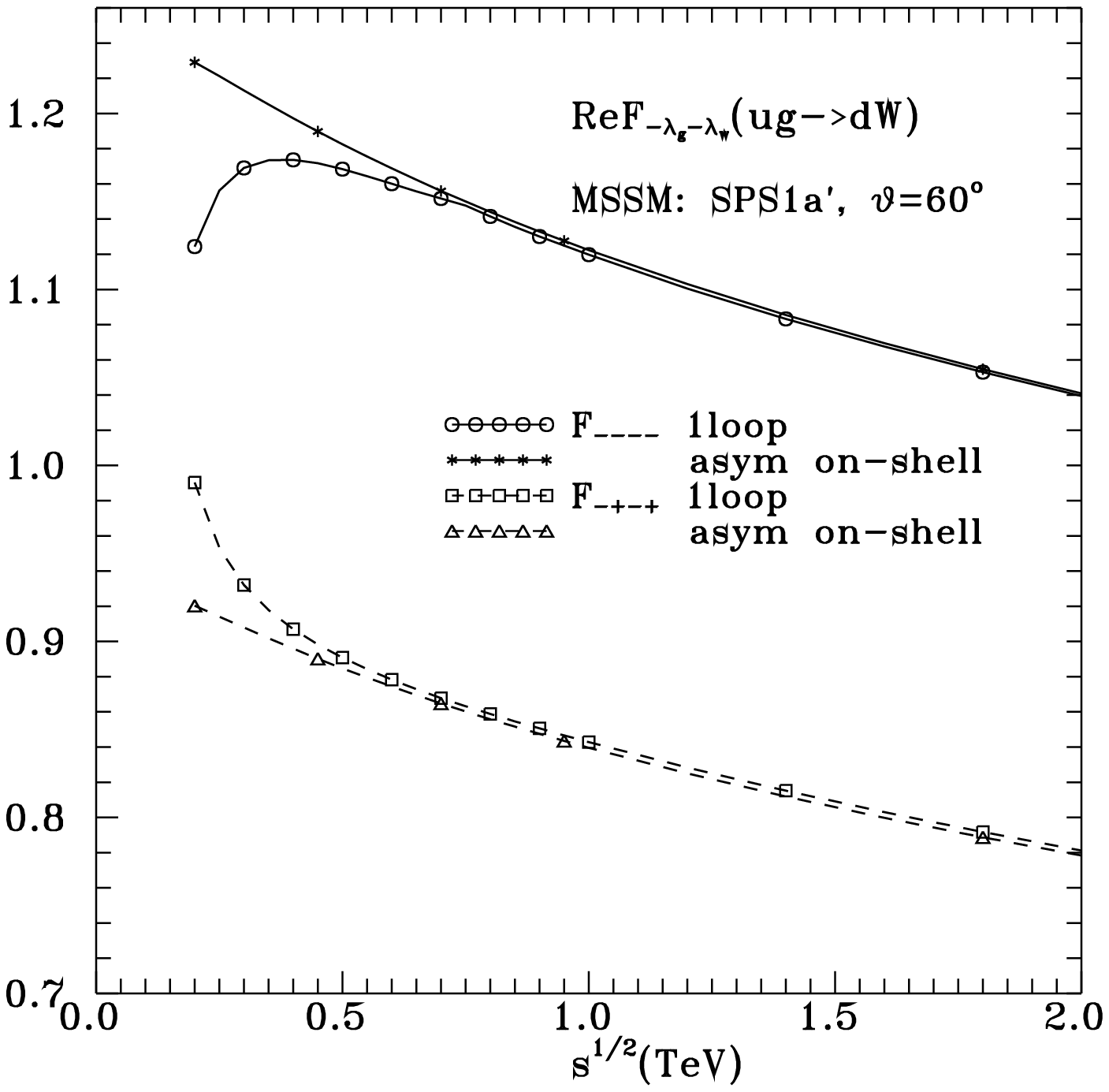, height=7.cm}\hspace{1.cm}
\epsfig{file=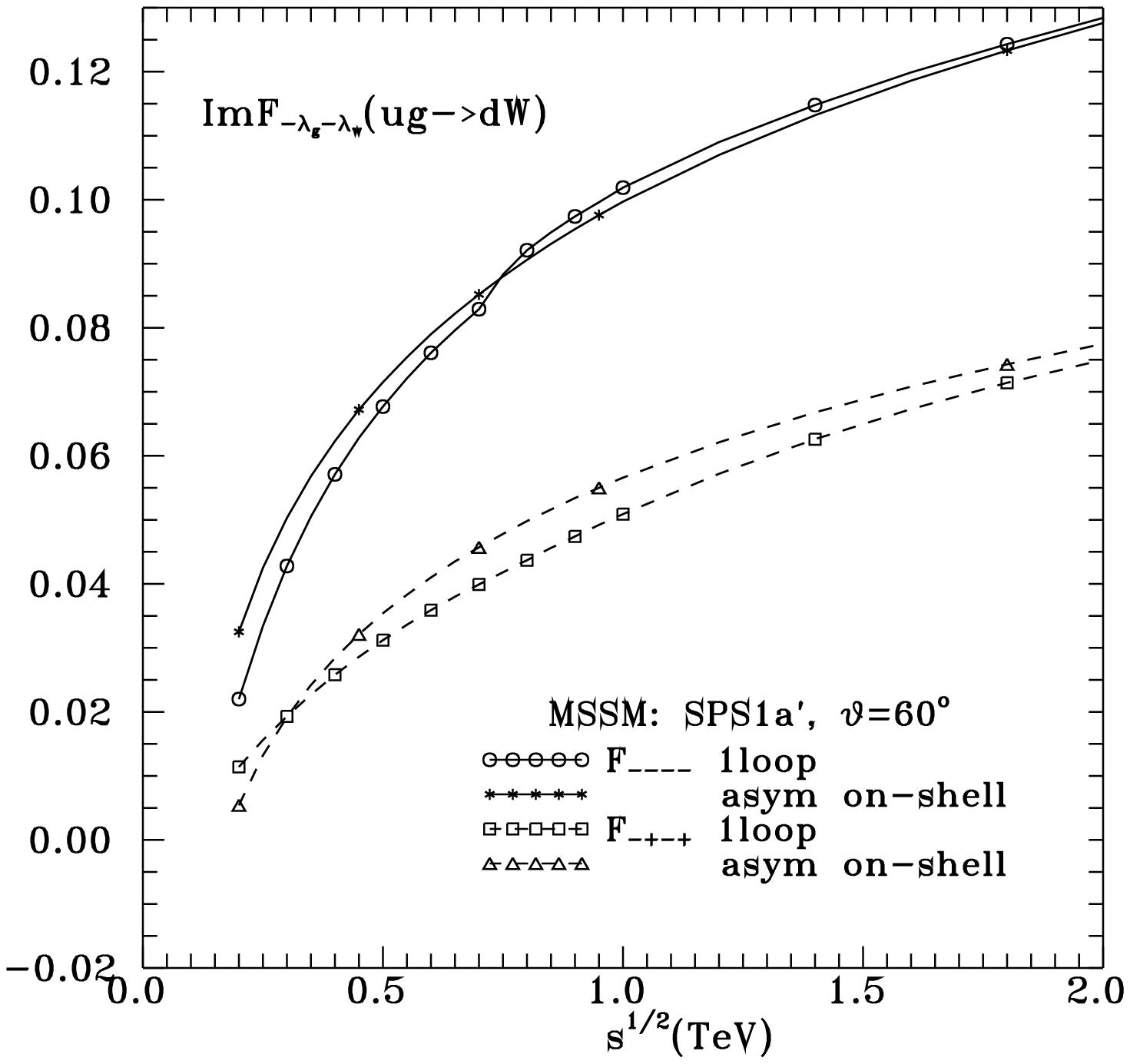,height=7.cm}
\]
\vspace*{0.1cm}
\[
\hspace{-0.5cm}
\epsfig{file=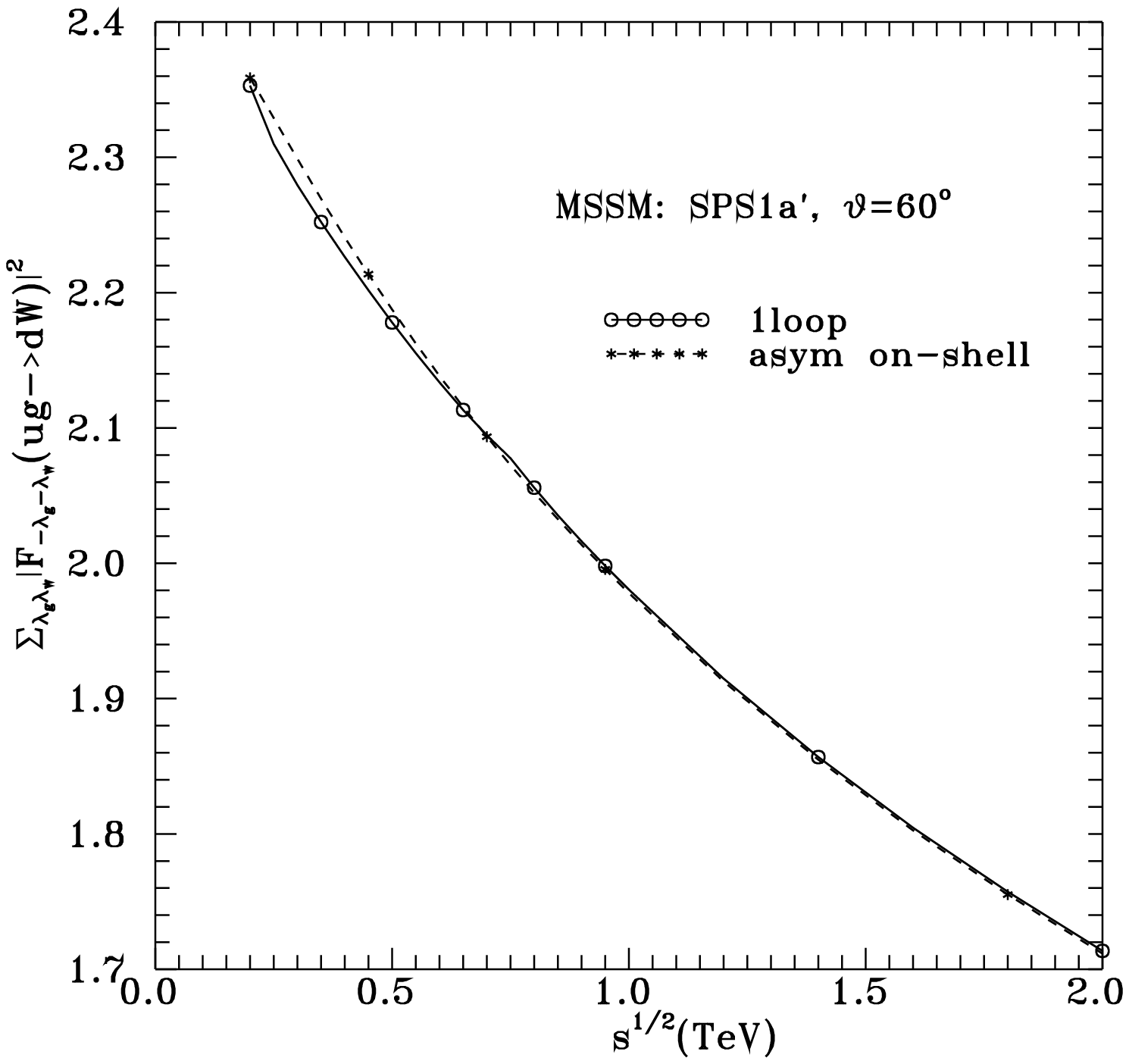, height=7.cm}\hspace{1.cm}
\epsfig{file=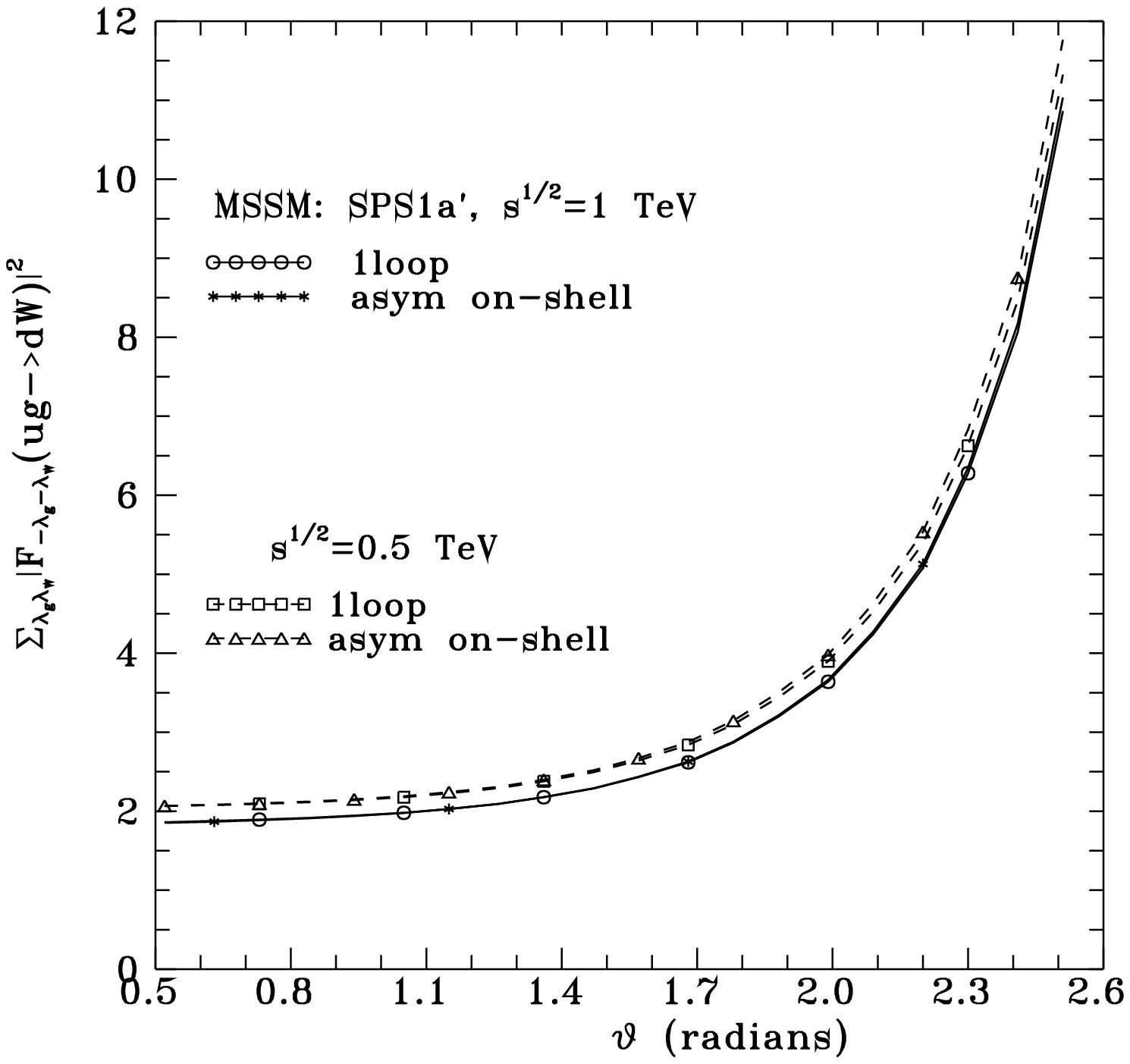,height=7.cm}
\]
\caption[1]{The complete 1loop results for $ug\to dW$ in $SPS1a'$
at the on-shell scheme \cite{ugdW},  are compared
to their high energy "supersimplicity" approximation.
Upper panels: Energy dependence of Real (left) and Im (right) parts of the HC amplitudes
  $F_{----}$ and $F_{-+-+}$ at $\theta=60^o$.
Lower panels: Sum over all amplitudes squared;
 energy  dependence  at $\theta=60^o$ (left);   angular dependence (right).  }
\label{SPS1ap-fig}
\end{figure}

\begin{figure}[p]
\vspace*{-1cm}
\[
\epsfig{file=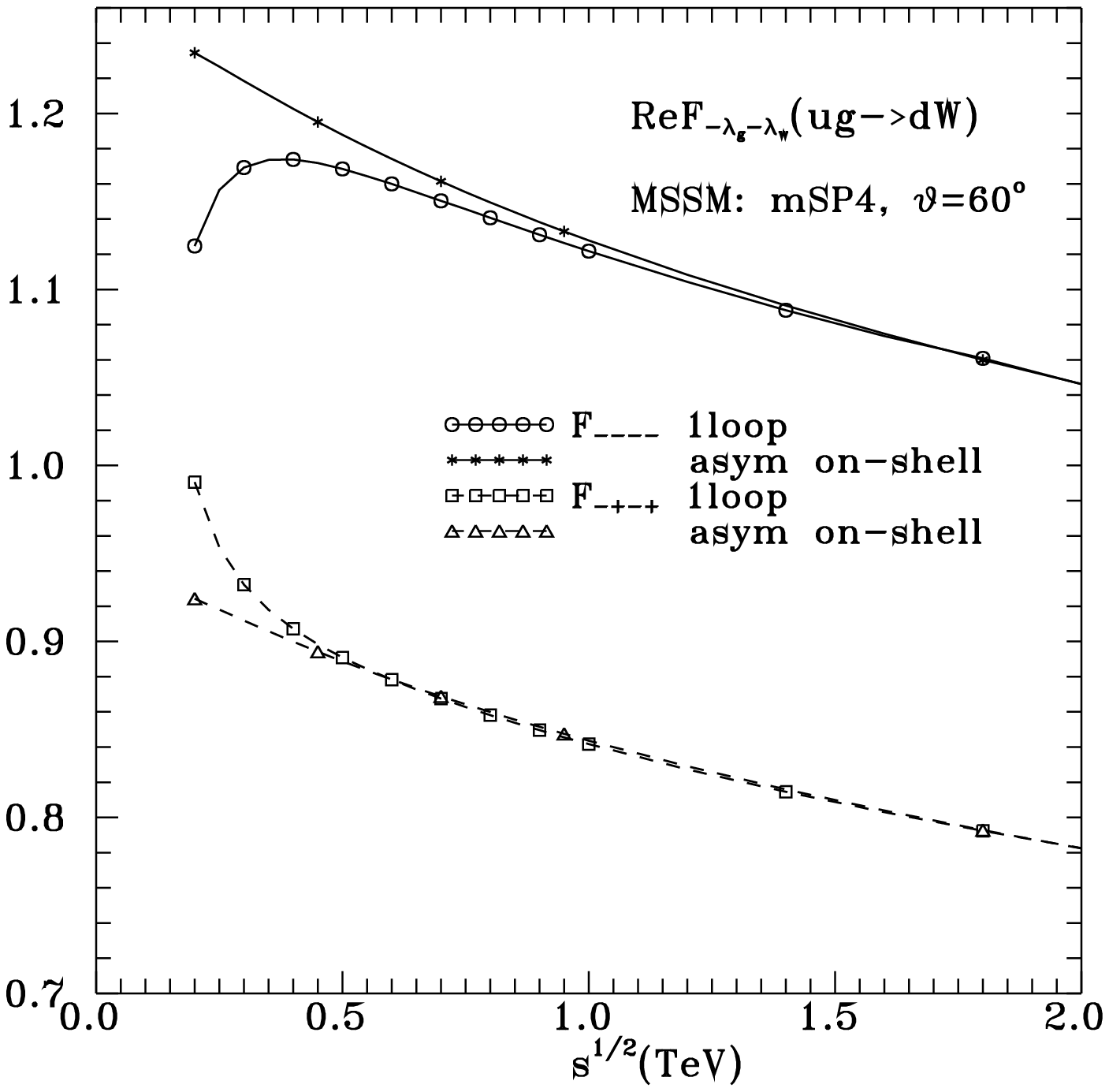, height=7.cm}\hspace{1.cm}
\epsfig{file=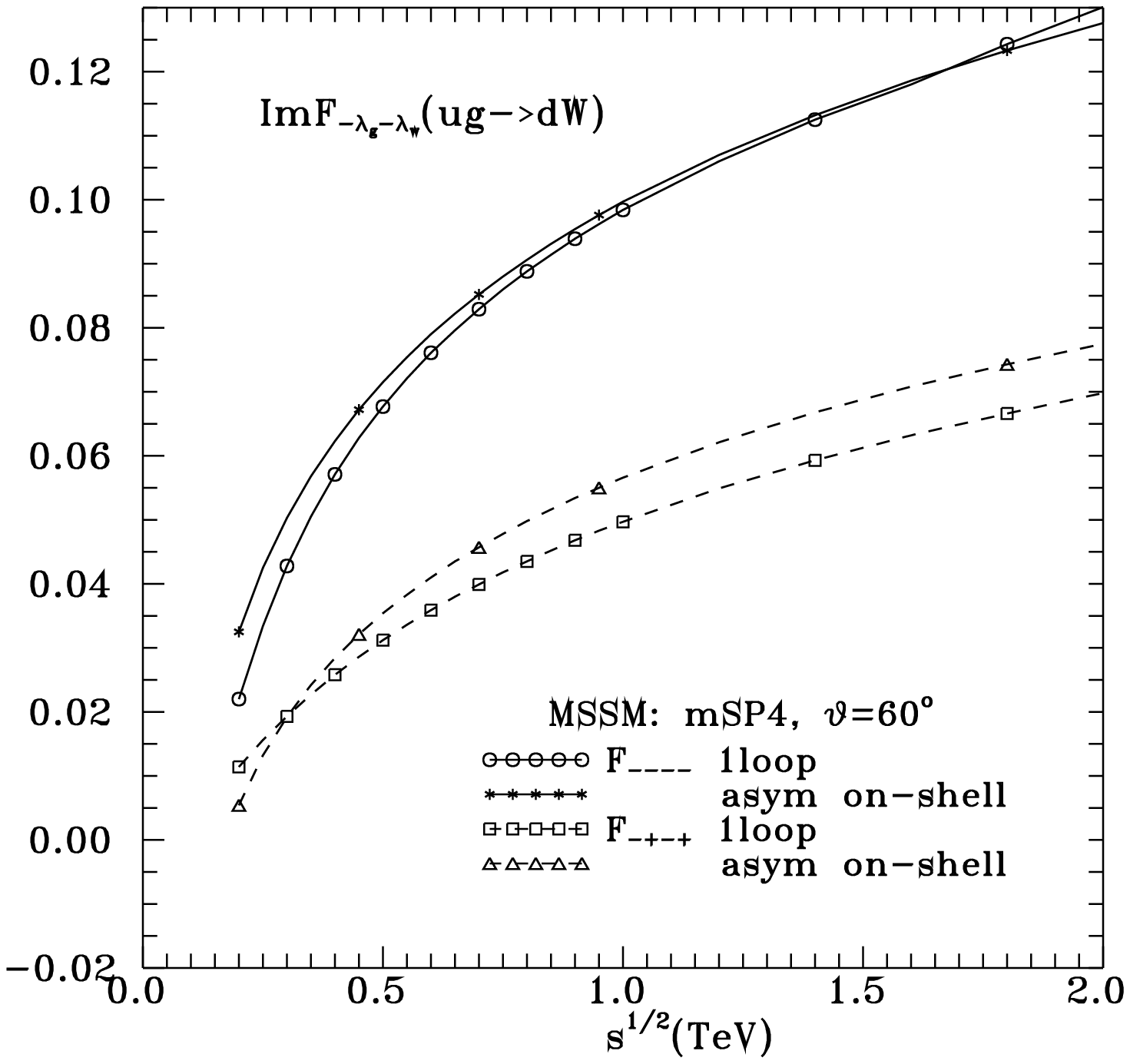,height=7.cm}
\]
\vspace*{0.1cm}
\[
\hspace{-0.5cm}
\epsfig{file=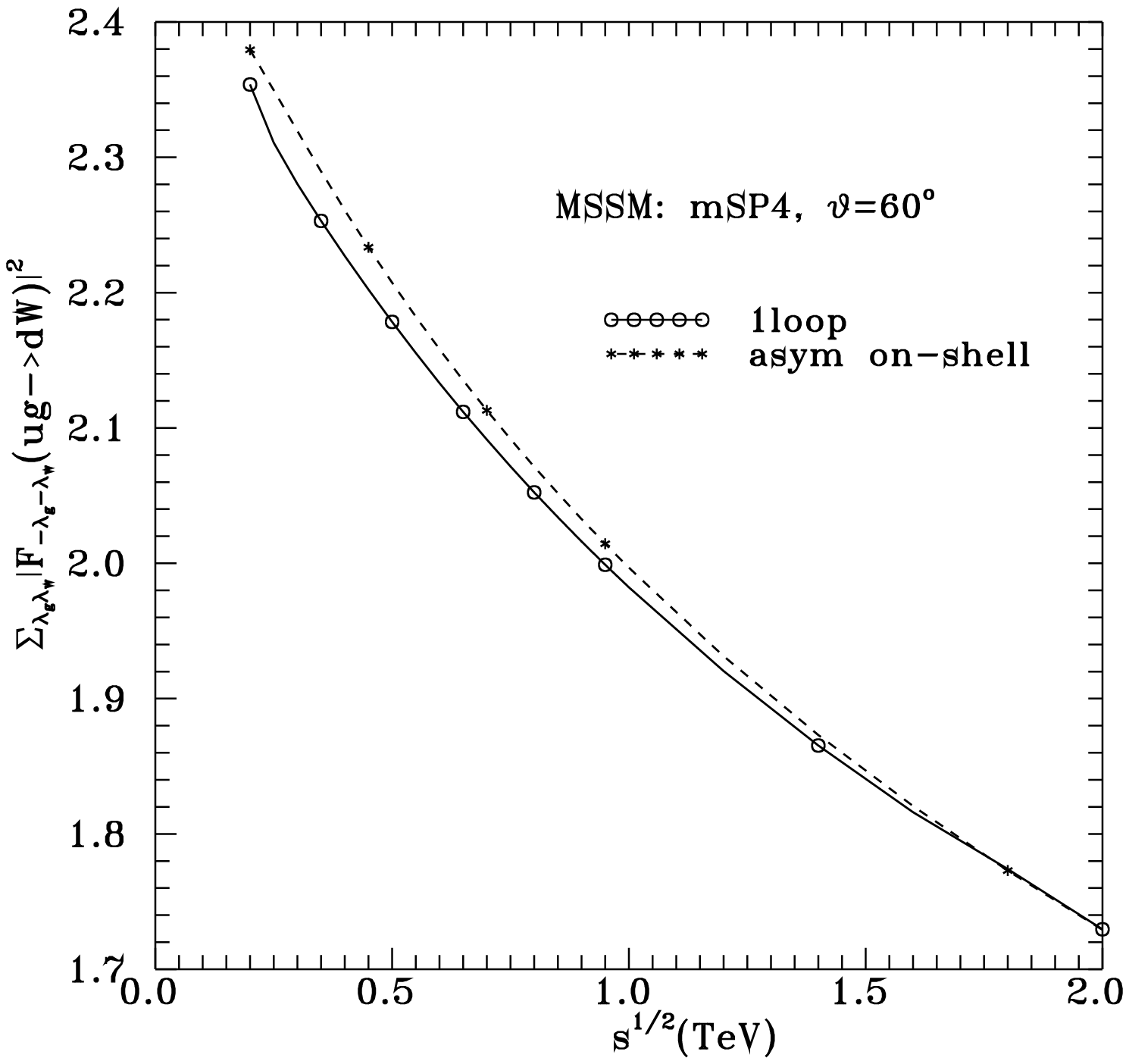, height=7.cm}\hspace{1.cm}
\epsfig{file=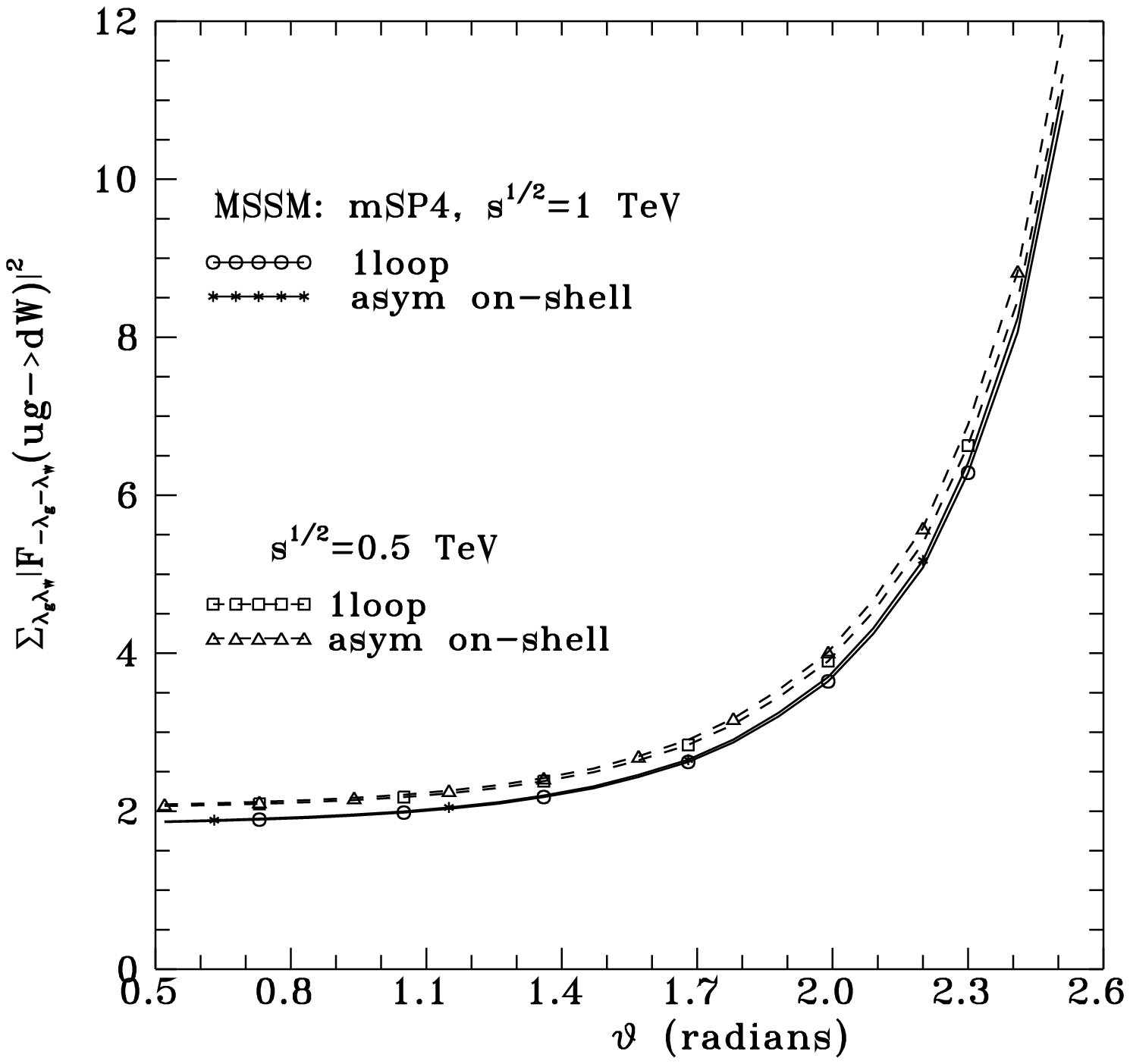,height=7.cm}
\]
\caption[1]{mSP4-results as in Fig.\ref{SPS1ap-fig}.  }
\label{mSP4-fig}
\end{figure}

\begin{figure}[p]
\vspace*{-1cm}
\[
\epsfig{file=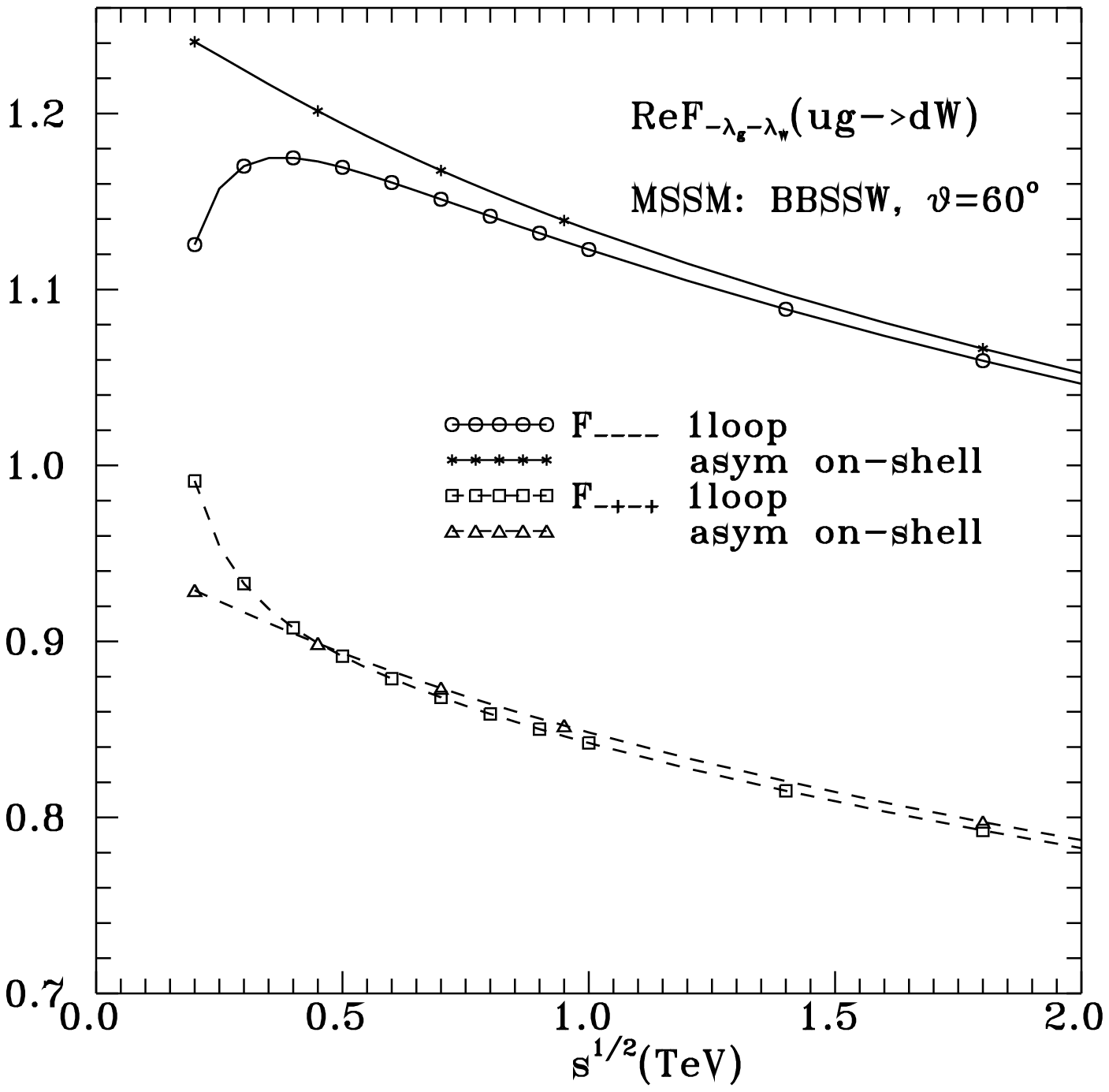, height=7.cm}\hspace{1.cm}
\epsfig{file=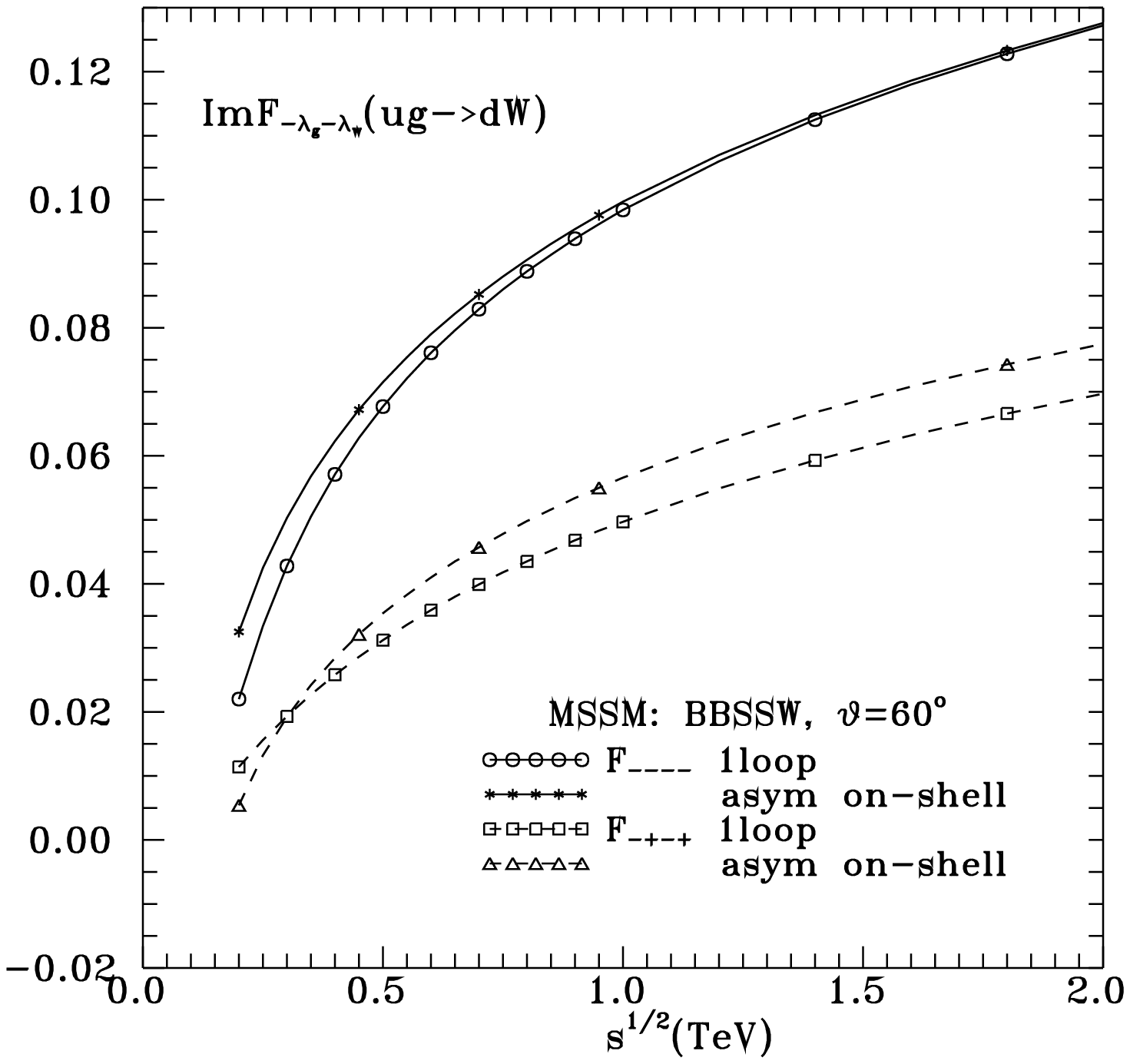,height=7.cm}
\]
\vspace*{0.1cm}
\[
\hspace{-0.5cm}
\epsfig{file=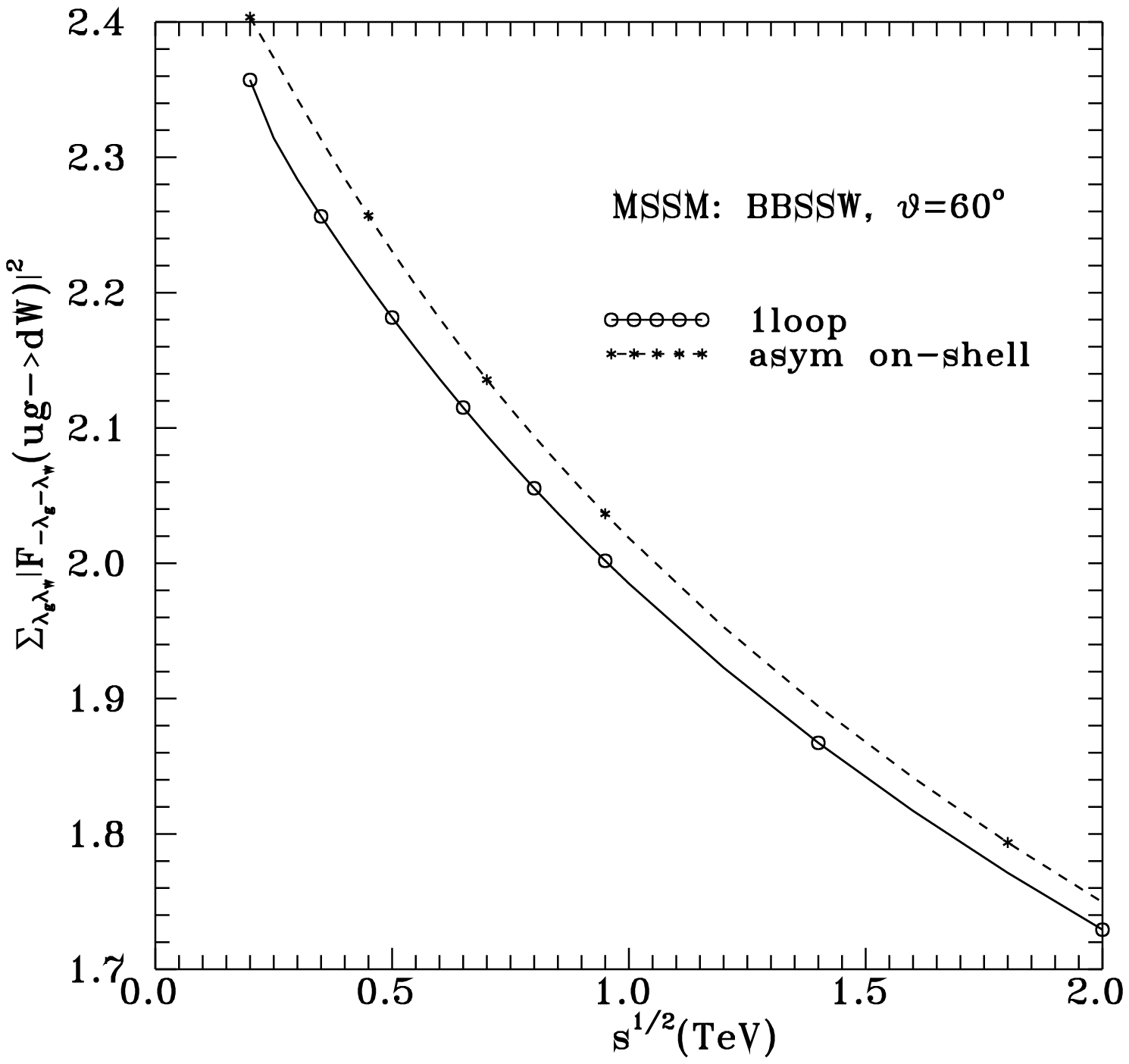, height=7.cm}\hspace{1.cm}
\epsfig{file=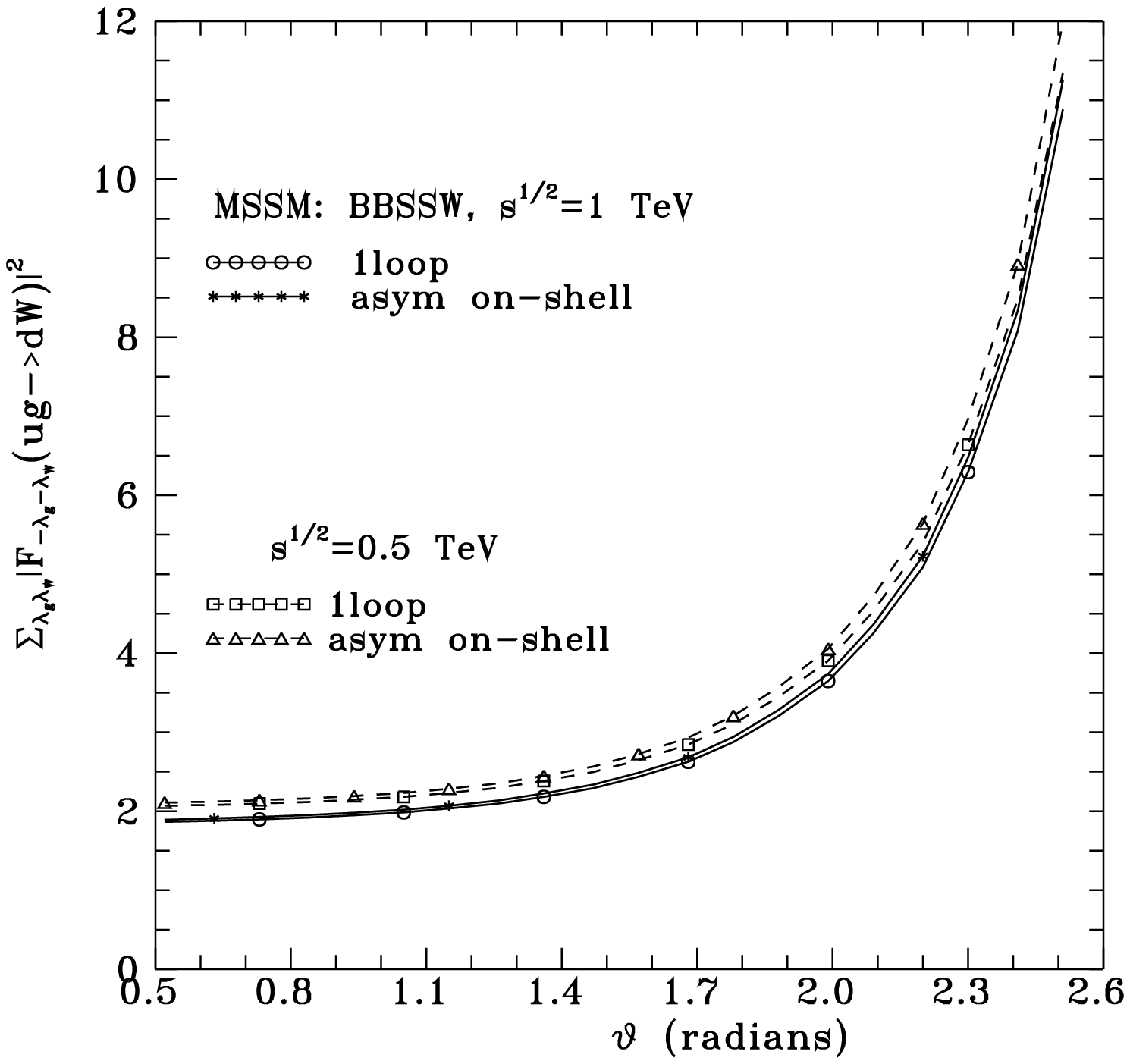,height=7.cm}
\]
\caption[1]{BBSSW-results as in Fig.\ref{SPS1ap-fig}.  }
\label{BBSSW-fig}
\end{figure}

\begin{figure}[h]
\vspace*{-0.5cm}
\[
\epsfig{file=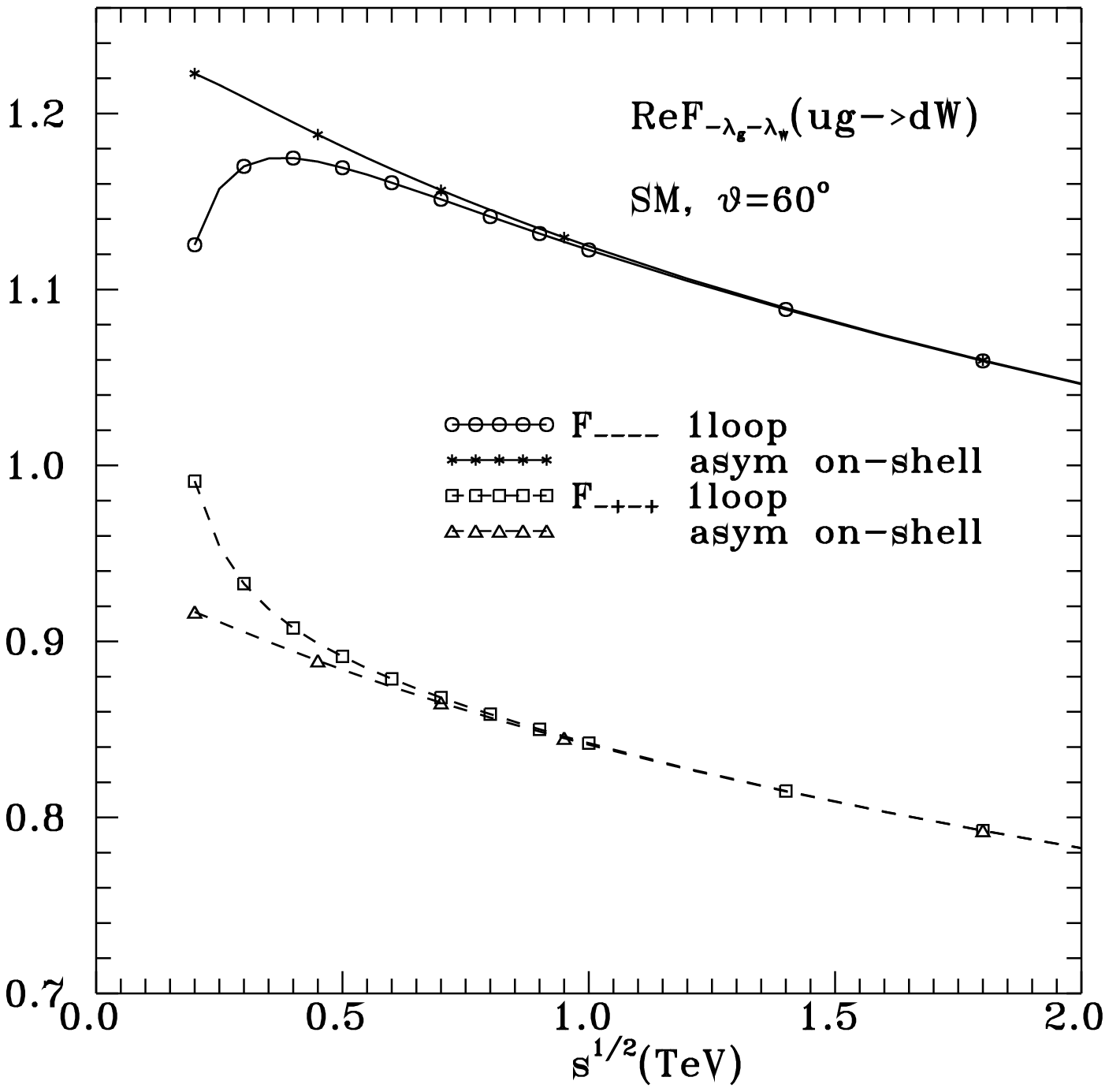, height=7.cm}\hspace{1.cm}
\epsfig{file=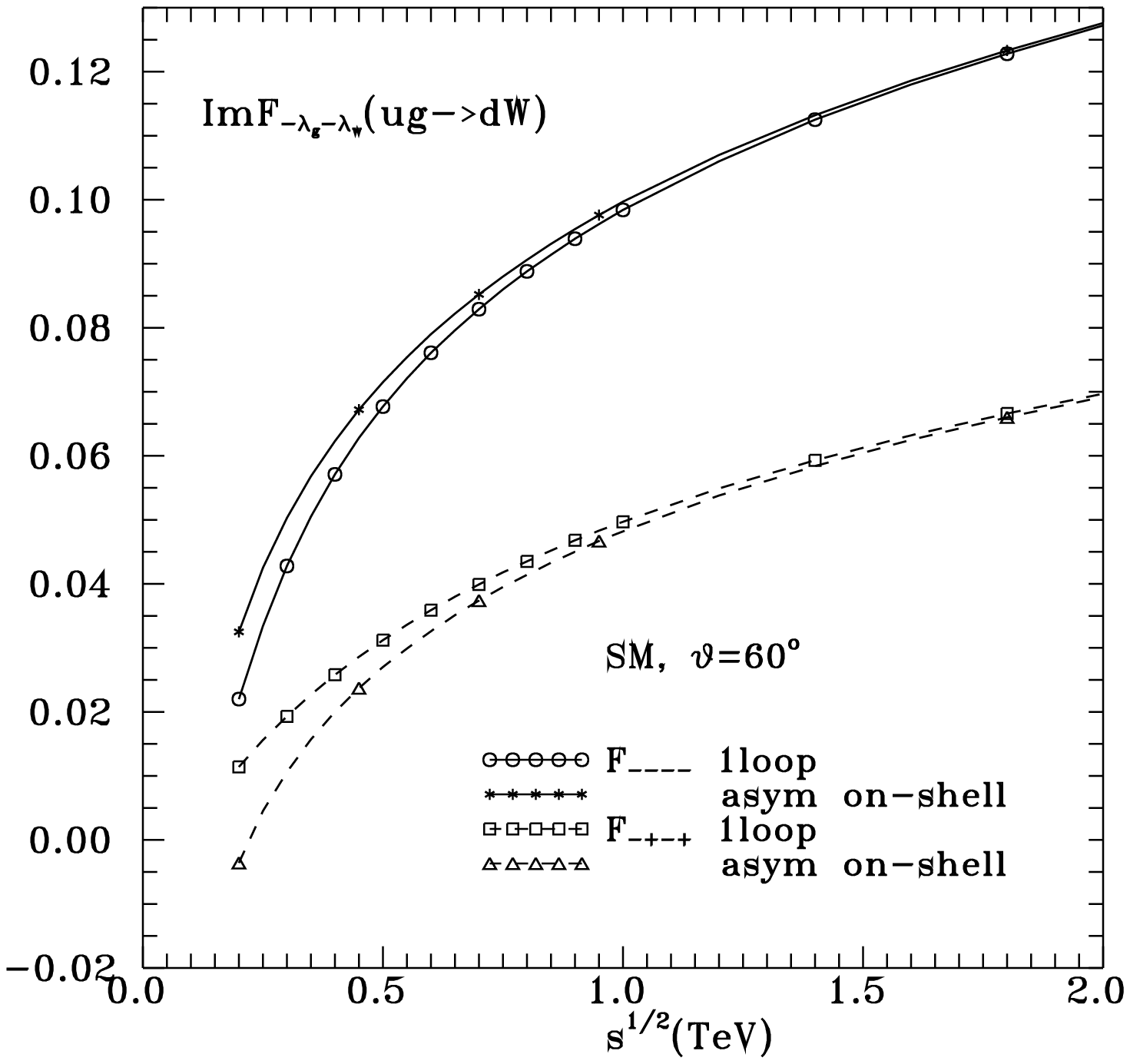,height=7.cm}
\]
\vspace*{0.1cm}
\[
\hspace{-0.5cm}
\epsfig{file=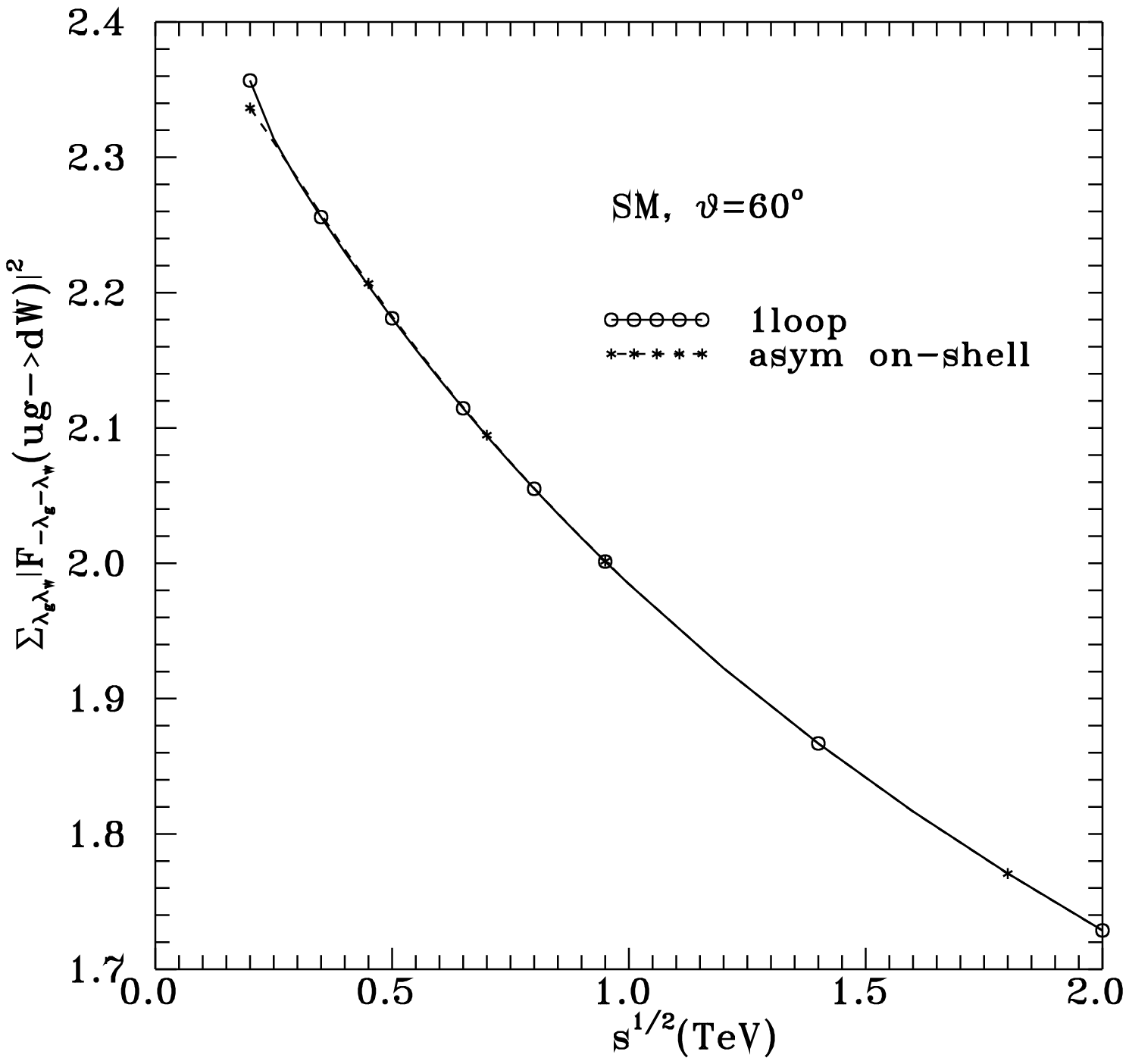, height=7.cm}\hspace{1.cm}
\epsfig{file=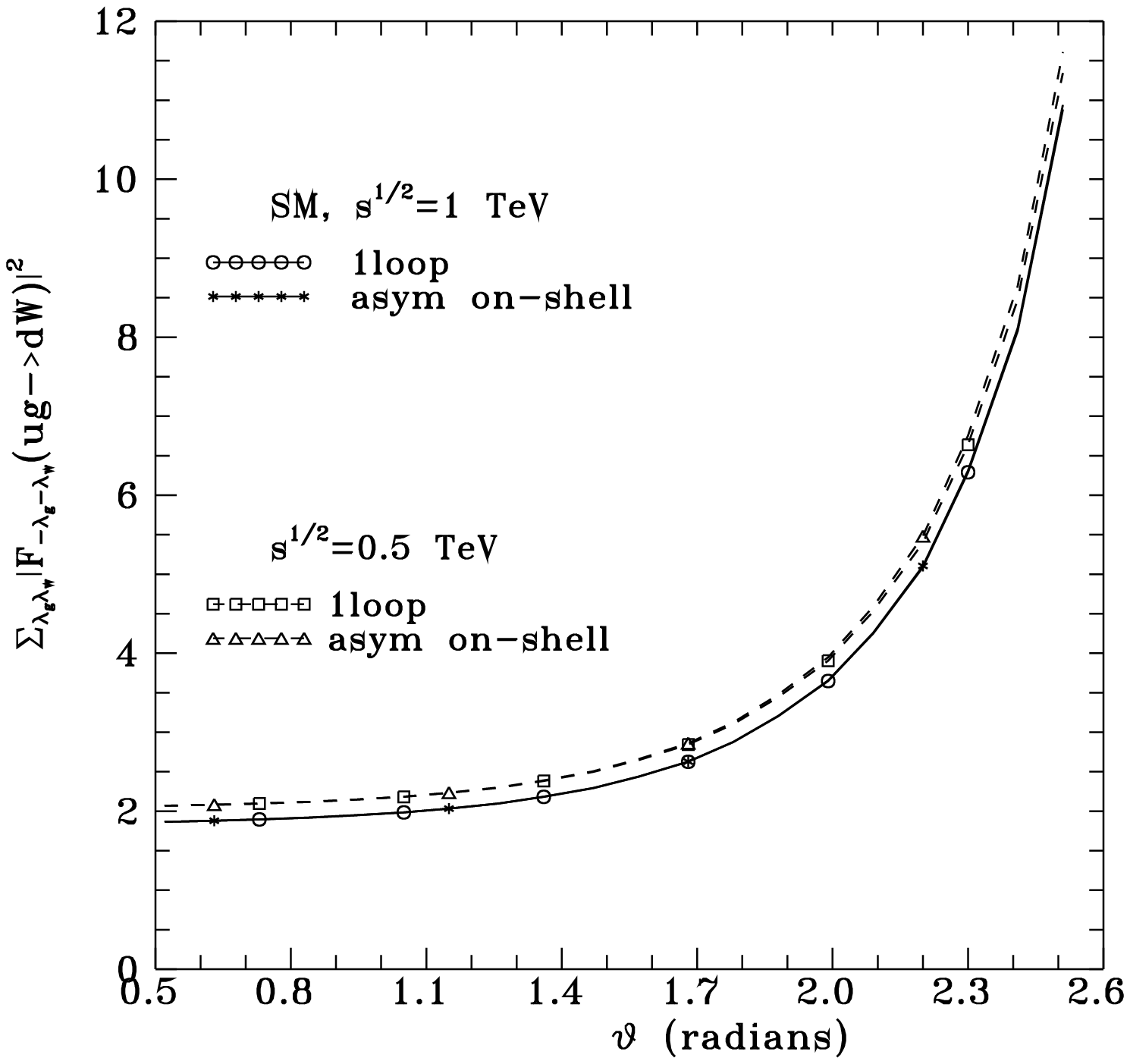,height=7.cm}
\]
\caption[1]{SM-results as in Fig.\ref{SPS1ap-fig}.  }
\label{SM-fig}
\end{figure}

Thus, in  Figs.\ref{SPS1ap-fig}, \ref{mSP4-fig}, \ref{BBSSW-fig}, we show the
HC amplitudes and the sum over amplitudes-squared
\[
\sum_{\lambda \mu \tau \mu'}|F_{\lambda \mu \tau \mu'}|^2
\]
for the  MSSM benchmarks in the first three lines of Table 1,
while in Figs.\ref{SM-fig},  the analogous results for SM are given.

As seen in Figs.\ref{SPS1ap-fig}, the high energy  {\it supersimplicity} structure  is
rather quickly established for $SPS1a'$ \cite{SPA1}.

In contrast, Figs.\ref{mSP4-fig}, \ref{BBSSW-fig}
indicate a much slower supersimplicity approach, for  the mSP4 \cite{mSP4}
and BBSSW \cite{Baer1} benchmarks, induced
by a considerably bigger SUSY scale; compare Table 1.
This seems stronger  for the imaginary parts of the amplitudes,
which are more sensitive to virtual thresholds. In any case the effect lies at  the 1\% percent
level, which  could be observable.

Corresponding results for  SM  are shown in Fig.\ref{SM-fig}.

In the lower right parts of all these figures, the angular distributions of
the exact 1loop and the asymptotic  expressions (\ref{F-asym-ugdw}), are compared.
As seen there, they roughly
  agree, already at 0.5TeV and  a wide range of angles, not only in SM, but also  for   the
   MSSM benchmarks \cite{SPA1, mSP4, Baer1}, even though the SUSY
   scale reaches quite high values. \\

 These remarks suggests a possibly simpler way to compare theory
with future experimental data.
This could be done by using the supersimplicity SRS expressions of (\ref{F-asym-ugdw-SRS}),
combined with an arbitrary real constant describing the residual counter term needed
for describing the on-shell amplitudes. Only one experimental input,
at an arbitrary energy and angle, should then  be sufficient to
fix the theoretical result.
We can then get a feeling of the energy domain in which the supersimplicity expressions
constitute a good approximation.\\

On the basis of the preceding discussion, we conclude, that the
high energy supersimplicity expressions (\ref{MSSM-Cmpmp}, \ref{MSSM-Cmmmm}) for MSSM,
and (\ref{SM-Cmpmp}, \ref{SM-Cmmmm}) for SM,  may
adequately  describe $ug\to dW$  at LHC  energies. The great virtue of these expressions,
is that they are analytical and very simple. Provided therefore the SUSY scale is not too high,
they constitute an efficient instrument for identifying the physics responsible for
the various effects. Particularly in the MSSM case,  they help identifying
 what are the SUSY-mass-combinations that mostly influence the various LHC observables.
 If needed, the accuracy of these
 predictions may  be further increased by including the residual counter term corrections
 $\delta_{OS}$, given  in Table 1 and (\ref{residual-SM-num}). \\

\section{Summary and prospectives for further studies}

By studying a large number of 2-to-2 MSSM process, at the 1loop EW order,
we have found that a remarkably simple structure arises for the HC amplitudes, which are
the only surviving ones   at high energy.

At such  energies and apart from a "residual constant",
these   amplitudes involve at most three different forms;
namely (\ref{d-tilde}, \ref{Sud-ln2}, \ref{Sud-ln}), containing the
well known logarithmic terms \cite{MSSMrules1, MSSMrules2, MSSMrules3},
to which  definite constants are added. The identification of these constants,
which greatly increase the accuracy of the high energy predictions,
is the main contribution of this work.

The MSSM high energy physical amplitudes are then expressed as  linear combinations of
these forms, with coefficients being rational functions of the $s,t,u$ variables;
and occasionally an additional  residual constant.
We have called this  very simple structure  of the high energy MSSM amplitudes,
{\it supersimplicity}.

Analogous results are also true for the SM case though, where
four log-involving forms are needed and additional  constants
are  inevitable, for describing the HC amplitudes.  We should remember though that in SM,
the HV amplitudes may occasionally be important.\\

If the Born-contribution is non-vanishing,
a special renormalization scheme, called SRS,  can be consistently defined
in either   MSSM or SM, where the validity
of supersimplicity   becomes asymptotically exact,
at the 1loop EW level. By this we mean that the asymptotic SRS HC amplitudes
are expressed as a linear combinations of three (four) forms for MSSM (SM)
respectively, without any residual constants.
These SRS amplitudes are  related
to the usual on-shell scheme, by adding to it a Born-like
contribution multiplied by a real residual counter term, relating the two schemes.

 For $ug\to dW$, this residual counter term has been found very small,
 for a wide range of  MSSM benchmarks in Table 1  and for SM.
Thus, for this process  at least, the {\it supersimplicity} structure is very
accurate. For achieving this, a  very important role is played by
 the  constants added to the logarithms in the forms
(\ref{d-tilde}, \ref{Sud-ln2}, \ref{Sud-ln}), which greatly enhance the accuracy
of the previously known logarithmic results \cite{MSSMrules1, MSSMrules2, MSSMrules3}.
This can be seen in  Figs.\ref{SPS1ap-fig}, \ref{mSP4-fig}, \ref{BBSSW-fig}, \ref{SM-fig},
where the exact 1loop results \cite{ugdW} are compared to the
 on-shell asymptotic ones, for three  MSSM benchmarks with a wide range of input parameters
  and SM.  These results show that the supersimplicity expressions provide
  a good description even
at rather low energies, when the SUSY scale is in the one TeV range. Even if
the SUSY scale is higher, these expressions constitute a good approximation
at the percent level.\\

Concentrating on  MSSM, we emphasize that the {\it supersimplicity}
 description of the asymptotic
HC amplitudes in terms of the three forms (\ref{d-tilde}, \ref{Sud-ln2}, \ref{Sud-ln}),
is not only a property of the Born-term involving processes
(\ref{process-Born}, \ref{mirror-process-Born}); but
it has also been seen in  $gg\to VV'$,  where just the form (\ref{d-tilde}) suffices.

And  it is also valid for the much more complicated
processes  $\gamma \gamma \to \gamma \gamma$, $\gamma Z, ZZ$, whose asymptotic
HC amplitudes may be fully expressed in terms of the forms (\ref{d-tilde}, \ref{Sud-ln2}),
again without any additional constant \cite{gamgamgamgam, gamgamgamZ, gamgamZZ}.

We are planning to further  explore this  in other processes, trying to see if there are any
exceptions. At present, we have   partial
 results for the process $e^+e^-\to f\bar f$ and its SUSY transformed
  $\tilde{e}^+\tilde{e}^-\to \tilde{f}\tilde{\bar f}$, which are consistent with those
  presented here. \\

We repeat  that {\it supersimplicity} is an 1loop MSSM property, realized
at the high energy region, where the   SUSY breaking effects are
either  minimized (as in the processes of Sect.3), or vanish completely
(as for the no Born processes of Sect.2.).
Diagrammatically, its realization involves two  steps. First the establishment
of Helicity Conservation, which is due  to  SUSY cancelations between fermionic
and bosonic diagrams; and second the actual derivation  of {\it supersimplicity},
for the  helicity conserving amplitudes, which are the only ones
that survive asymptotically.

At the technical level,   the easiest way
to establish {\it supersimplicity} for processes involving
external gauge bosons\footnote{These are the processes
where HCns is most intriguing \cite{Corfu}. },
 is  to use their SUSY-transformed process,
find the asymptotic HC amplitudes
there,  and then transform back to the original process,
appropriately changing the internal and external masses in the forms
(\ref{Sud-ln}, \ref{Sud-ln2}). \\

In SM, there is no helicity conservation theorem in general.
Nevertheless, restricting to HC amplitudes, a corresponding analysis
may be made. The main result now is that an additional fourth form appears,
involving linear logarithms of ratios of Mandelstam variables; and additional
constants may be occasionally  needed.\\

In conclusion, we emphasize that  {\it supersimplicity}, describing
the leading HC amplitudes through formulae of a few lines in MSSM,
is appealing from two aspects. The first one is theoretical;  the simplicity of
these formulae allows one to immediately read what are the main high-energy  features
of the electroweak contributions to the process considered
and what is the role of supersymmetry.

The second one concerns the future comparison with experiments.
In this paper we have concentrated on the 1loop electroweak effects. A complete analysis
will of course require the additional treatment of the QED and QCD corrections
(in particular soft photon and gluon radiation) for which there exist specific codes.
For what concerns QCD in particular,  we just note that the  SUSY QCD  high energy
contribution should behave similarly to the EW gauge part effects  studied here.
Thus, the Sudakov logarithms in e.g.   \cite{stop} should be replaced
by  augmented forms similar to those  of Section 3 (replacing charginos by gluinos).
In addition,  a study of the relevant background processes
will also  be necessary for each case. Nevertheless, our modest work  may  be useful
in this respect also, since it could allow   someone to get a general feeling of the
high energy  effects, by using simple formulas like (\ref{MSSM-Cmpmp}, \ref{MSSM-Cmmmm}),
instead of the enormous    codes containing  the  exact 1loop EW  virtual
corrections.

\vspace*{0.5cm}
\noindent
{\bf Acknowledgements} We are grateful to Jacques Layssac for help in using
the $ug\to dW$ code. GJG is  partially supported by the European Union
 contracts MRTN-CT-2006-035505 HEPTOOLS  and ITN programme "UNILHC" PITN-GA-2009-237920.

\vspace*{1cm}
\appendix

\renewcommand{\thesection}{}
\renewcommand{\theequation}{A.\arabic{equation}}
\setcounter{equation}{0}

\section{Appendix:   High energy  structure  of    $bg\to bH^0_i$, \\ at 1loop EW. }

In analogy to the $ug\to dW$ analysis in Sect. 3.2,
we here present the high energy HC amplitudes
for the process $b g\to bH^0_i$, which is sensitive to the Higgs (Yukawa)
sector, in both MSSM and SM. Here
 $H^0_i=h^0, H^0, A^0, G^0$,  describes any of the neutral  Higgs
 or Goldstone bosons in MSSM; while in SM,  $H^0_i=H_{SM}, G^0$ .

 The helicity amplitudes for $bg\to bH^0_i$ are denoted
 by $F_{\lambda \mu \tau}$, with $(\lambda, \tau)$
describing the helicities of the initial and final $b$-quark, while  $\mu$ denotes the helicity
of the gluon. The asymptotic Born contributions to these processes are
\bq
F^{\rm Born}_{-++} = -\sqrt{2}c_{H^0_i}^Lg_s \, {t\over u}\cos{\theta\over2}
~~~,~~~ F^{Born}_{+--}= -\sqrt{2}c_{H^0_i}^Rg_s \, {t\over u} \cos{\theta\over2}~~~,
\label{bgbH-Born}
\eq
with the MSSM couplings being
\bqa
c^L_{H^0}=c^R_{H^0}=-~{em_b\cos\alpha\over2s_Wm_W\cos\beta} &~,~&
c^L_{h^0}=c^R_{h^0}=~{em_b\sin\alpha\over2s_Wm_W\cos\beta}~~, \nonumber\\
c^L_{A^0}=-c^R_{A^0}=-i{em_b\tan\beta\over2s_Wm_W} &~,~&
c^L_{G^0}=-c^R_{G^0}=i{em_b\over2s_Wm_W} ~~~. \label{bbH-cs}
\eqa

Using the SUSY transformed process
$\tilde{b}\tilde{g}\to \tilde{b}\tchi^0_i$ for simplifying the calculations, and
 selecting the higgsino components, one gets in the SRS scheme in MSSM
\bq
F_{\mp\pm\pm}^{\rm SRS}=F^{\rm Born}_{\mp\pm\pm}
\left [1+{\alpha\over4\pi}C_{\mp\pm\pm}(s,t,u) \right]~~, \label{F-asym-bgbH-SRS}
\eq
where
\bqa
 C_{+--}^{\rm MSSM}(s,t,u)& = &C_{-++}^{\rm MSSM}(u,t,s)=
- {1\over18c^2_W}[-\overline{\ln^2t_Z} +(\ln^2r_{ts}+\pi^2)+ (\ln^2r_{tu}+\pi^2) ]
\nonumber\\
&&- {(1+2c^2_W) \over 12s^2_Wc^2_W}\, \overline{\ln^2s_Z}
- {(1+8c^2_W) \over 12s^2_Wc^2_W}\, {s\over t}(\ln^2 r_{us}+\pi^2) \nonumber\\
&&
+{1\over 6c^2_W}[-\overline{\ln^2u_Z} - {u\over t}(\ln^2r_{us} +\pi^2) ]~~.
\label{MSSM-C-bgbH}
\eqa
 Thus, in MSSM,  only the forms  (\ref{d-tilde}) and (\ref{Sud-ln2}) appear,
while no Sudakov linear log forms, like those defined in (\ref{Sud-ln}), arise.
Note that (\ref{MSSM-C-bgbH}) holds the same for all $H^0_i$, since no couplings
 like those in (\ref{bbH-cs}) appear in it.

The   needed Sudakov $\ln^2$ forms in (\ref{MSSM-C-bgbH}),  are
\bqa
\overline{\ln^2 t_Z} &=& \ln^2 \frac{-t-i\epsilon}{\mzsq}+4 L_{bZb} ~~, \nonumber \\
\overline{\ln^2 s_Z} &=& \ln^2 \frac{-s-i\epsilon}{\mzsq}+2( L_{H^0_iZ\varphi^0}+L_{bZb})
~~, \nonumber \\
\overline{\ln^2 s_W} &=& \ln^2 \frac{-s-i\epsilon}{\mwsq}+2( L_{bWt}+L_{H^0_iW\varphi^-})
~~, \nonumber \\
\overline{\ln^2 u_Z} &=& \ln^2 \frac{-u-i\epsilon}{\mzsq}+2( L_{H^0_iZ\varphi^0}+L_{bZb})
~~,  \label{Sud-ln2-bgbH}
\eqa
where $\varphi^0, \varphi^-$ respectively describe  mixtures of the  Higgs
or Goldstone internal lines in the contributing diagram.
Together with the corresponding   $V$-internal lines, these
generate the  terms $L_{H^0_iZ\varphi^0}, ~L_{H^0_iW \varphi^-}$,
contributing to the  $H^0_i$ production.  The explicit meanings of these terms are
\bqa
L_{H^0Z\varphi^0} & =&  {\sin\beta \sin(\beta-\alpha)\over\cos\alpha}L_{H^0ZA^0}+
{\cos\beta \cos(\beta-\alpha)\over\cos\alpha}L_{H^0ZG^0}~~, \nonumber \\
L_{H^0W\varphi^-} & =&{\sin\beta \sin(\beta-\alpha)\over\cos\alpha}L_{H^0WH^-}+
{\cos\beta \cos(\beta-\alpha)\over\cos\alpha}L_{H^0WG^-}~~,  \nonumber \\
L_{h^0Z\varphi^0} & =&{\sin\beta \cos(\beta-\alpha)\over\sin\alpha}L_{h^0ZA^0}-
{\cos\beta \sin(\beta-\alpha)\over\sin\alpha}L_{h^0ZG^0}~~, \nonumber \\
L_{h^0W\varphi^-} & =& {\sin\beta \cos(\beta-\alpha)\over\sin\alpha}L_{h^0WH^-}-
{\cos\beta \sin(\beta-\alpha)\over\sin\alpha}L_{h^0WG^-}~~, \nonumber \\
L_{A^0Z\varphi^0} & =& {\cos\alpha \sin(\beta-\alpha)\over\sin\beta}L_{A^0ZH^0}+
{\sin\alpha \cos(\beta-\alpha)\over\sin\beta}L_{A^0Zh^0}~~,\nonumber \\
L_{A^0W\varphi^-} & =& L_{A^0WH^-} ~~,\nonumber \\
L_{G^0Z\varphi^0} & =& {\cos\alpha \cos(\beta-\alpha)\over\cos\beta}L_{G^0ZH^0}-
{\sin\alpha \sin(\beta-\alpha)\over\cos\beta}L_{G^0Zh^0}~~, \nonumber \\
L_{G^0W\varphi^-} & =& L_{G^0WG^-} ~~, \label{Sud-ln2-bgbH1}
\eqa
where  (\ref{LaVc-term}) should be used.
Notice that in the r.h.s. of  all equations  (\ref{Sud-ln2-bgbH1}),
 the sum of the coefficients of the  $L_{abc}$ forms  equals to 1,
 as it should.

As we have already said, the results (\ref{F-asym-bgbH-SRS},\ref{MSSM-C-bgbH})
 were derived by working with the process
$\tilde{b}\tilde{g}\to \tilde{b}\tchi^0_i$, and  their
logarithmic behavior in (\ref{F-asym-bgbH-SRS},\ref{MSSM-C-bgbH}) should agree
 with the old  Sudakov structure established directly for $bg\to bH^0_i$  \cite{MSSMrules3}.
As has amply been pointed out above, the absence of linear logs in (\ref{MSSM-C-bgbH}),
 is  an  MSSM feature.\\

To check what happens in the SM cases,  a direct diagrammatic computation must
be made.

For  $H^0_i = H_{SM}$, we would then use $L_{H_{SM}ZG^0}$, $L_{H_{SM}WG^-}$
in (\ref{Sud-ln2-bgbH}) and the couplings
\bq
c^L_{H^0_{SM}}=c^R_{H^0_{SM}}=-{em_b \over 2s_Wm_W} ~~ ~. \label{bbH-cs-SM}
\eq
Compared to the  MSSM expressions (\ref{F-asym-bgbH-SRS},\ref{MSSM-C-bgbH}),
the 1loop SM correction  contains  typical  linear terms  $\ln r_{us}$ ~,
together with contributions of  the forms (\ref{d-tilde}, \ref{Sud-ln}).  We find
for  $H^0_i = H_{SM}$,
\bqa
C^{\rm SM}_{\mp\pm\pm} -C^{\rm MSSM}_{\mp\pm\pm} &=&{1+2c^2_W\over2s^2_Wc^2_W}
\left [ -{su\over2t^2} \left (\overline{\ln^2 r_{us}}+\pi^2 \right )
+{u\over t}\ln r_{us} \right ]\nonumber\\
&&
+{\overline{\ln u_{ZG^0}} \over2s^2_Wc^2_W}
+{\overline{\ln u_{WG}} \over s^2_W}
-{m^2_t\over2s^2_W m^2_W} \left [\overline{\ln u_{tG}} +{u\over t}\ln r_{us}\right ] ~~,
\label{SM-MSSM-C-bgbH}
\eqa
where $G\equiv G^{\pm}$ denotes a charged Goldstone boson.

For the case $H^0_i=G^0$ in SM, one should use $L_{G^0ZH_{SM}},L_{G^0WG^-}$, leading
\bqa
C^{\rm SM}_{\mp\pm\pm} -C^{\rm MSSM}_{\mp\pm\pm} &=&{1+2c^2_W\over2s^2_Wc^2_W}
\left [ -{su\over2t^2} \left (\overline{\ln^2 r_{us}}+\pi^2 \right )
+{u\over t}\ln r_{us} \right ]\nonumber\\
&&
+{\overline{\ln u_{ZH_{SM}}} \over2s^2_Wc^2_W}
+{\overline{\ln u_{WG}} \over s^2_W}
-{m^2_t\over2s^2_W m^2_W} \left [\overline{\ln u_{tG}} +{u\over t}\ln r_{us}\right ] ~~.
\label{SM-MSSM-C-bgbG}
\eqa

In  (\ref{SM-MSSM-C-bgbH}, \ref{SM-MSSM-C-bgbG}) as well as in (\ref{MSSM-C-bgbH}),
only  contributions of the  {\it supersimplicity} structure arise,
containing the  forms (\ref{d-tilde}, \ref{Sud-ln2}, \ref{Sud-ln}) and
linear logarithms of ratios of the $s,t,u$ variables,
appear. Therefore, these are the HC amplitudes in the SRS scheme.
To find the on-shell amplitudes,
the counter term  contributions, analogous to (\ref{residual-COS}),
 must be calculated. This  has not been done here.

\end{document}